\documentclass{emulateapj}

\usepackage[]{amsmath}
\usepackage{amssymb}
\usepackage{natbib}
\usepackage{graphicx}
\usepackage{multirow}
\usepackage{booktabs}
\usepackage{bm}
\usepackage{xspace}
\usepackage{afterpage}
\usepackage{gensymb}
\usepackage{newtxtext,newtxmath}
\usepackage[T1]{fontenc}
\usepackage{ae,aecompl}
\usepackage{booktabs} 
\usepackage[dvipsnames]{xcolor}

\usepackage{color}			      
\definecolor{midgray}{gray}{0.4}		
\definecolor{orange}{rgb}{1,0.5,0}    
\usepackage[colorlinks=true,citecolor=blue,linkcolor=midgray,urlcolor=midgray]{hyperref}        

\slugcomment{Accepted for Publication in MNRAS, September 2017}
\shorttitle{Pathways for compact elliptical galaxy formation}
\shortauthors{A. Ferr\'e-Mateu et al.}

\begin{document}

\title{On the formation mechanisms of compact elliptical galaxies}
\author{Anna Ferr\'e-Mateu \altaffilmark{1}}
\author{Duncan A. Forbes \altaffilmark{1}}
\author{Aaron J. Romanowsky \altaffilmark{2,3}}
\author{Joachim Janz \altaffilmark{1,4,5}}
\author{Christopher Dixon \altaffilmark{2}} 
\affil{\altaffilmark{1}Centre for Astrophysics \& Supercomputing, Swinburne University of Technology, Hawthorn VIC 3122, Australia}
\affil{\altaffilmark{2}Department of Physics and Astronomy, San Jos\'e State University, San Jose, CA 95192, USA }
\affil{\altaffilmark{3}University of California Observatories, 1156 High St., Santa Cruz, CA 95064, USA}
\affil{\altaffilmark{4}Astronomy Research Unit, University of Oulu, FI-90014 Finland}
\affil{\altaffilmark{5}Finnish Centre of Astronomy with ESO (FINCA), University of Turku, Vaisalantie 20, 21500 Piikkio, Finland }
\email{aferremateu@swin.edu.au (AFM)}   

\begin{abstract}
In order to investigate the formation mechanisms of the rare compact elliptical galaxies (cE) we have compiled a sample of 25 cEs with good SDSS spectra, covering a range of stellar masses, sizes and environments. They have been visually classified according to the interaction with their host, representing different evolutionary stages. We have included clearly disrupted galaxies, galaxies that despite not showing signs of interaction are located close to a massive neighbour (thus are good candidates for a stripping process), and cEs with no host nearby. For the latter, tidal stripping is less likely to have happened and instead they could simply represent the very low-mass, faint end of the ellipticals. We study a set of properties (structural parameters, stellar populations, star formation histories and mass ratios) that can be used to discriminate between an intrinsic or stripped origin. We find that one diagnostic tool alone is inconclusive for the majority of objects. However, if we combine all the tools a clear picture emerges. The most plausible origin, as well as the evolutionary stage and progenitor type, can be then determined. Our results favor the stripping mechanism for those galaxies in groups and clusters that have a plausible host nearby, but favors an intrinsic origin for those rare cEs without a plausible host and that are located in looser environments.
\end{abstract}



\section{Introduction}
Compact ellipticals (cEs) are relatively rare galaxies in the local Universe, with only about two hundred objects currently known (\citealt{Chilingarian2009};  \citealt{Norris2014}; \citealt{Chilingarian2015}). They are non star forming galaxies typically characterized by low stellar masses of 10$^{8} \la $M$_{*}$/M$_{\odot} \la $10$^{10}$, very compact effective radii of 100 $\la \mathrm{R_e} \la$ 1000\,pc and very high stellar densities. The high stellar densities inferred for cEs are similar to those in the cores of early-type galaxies (ETGs) or the bulges of spirals, suggesting that cEs could be the remnant cores of larger galaxies that have been tidally stripped of their outer stars. This claim is also supported by fact that they seem to follow as the low-mass, low-luminosity extension described by bright and massive ellipticals, branching off the well-known luminosity- and mass-size relations (e.g. \citealt{Brodie2011}; \citealt{Misgeld2011}). This distinctively differentiates them from their low density counterparts dwarf ellipticals (dE; \citealt{Chilingarian2009}). In fact, the vast majority of known cEs are located close to a much larger host galaxy, which plausibly has caused the threshing. In addition, they are preferentially located in clusters or populous groups. The smoking gun for this scenario is that a small number of cEs have been `caught in the act', interacting with their host (e.g. \citealt{Huxor2011}; \citealt{Paudel2013}; \citealt{Chilingarian2015}). However, this mechanism has been challenged by the discovery of a few isolated cEs. Although some of these isolated cEs are compatible with being galaxies that have been ejected from the environments where they were originally stripped \citep{Chilingarian2015}, other galaxies in such group demand an alternative mechanism of formation (e.g. \citealt{Wirth1984}; \citealt{Huxor2013}; \citealt{Paudel2014b}; \citealt{Wellons2016}).

Therefore, different formation mechanisms seem to be plausible to create the family of cEs. In fact, two main mechanisms have been largely debated, similarly to the mechanisms that form ultra compact dwarfs (UCDs).  The first mechanism is tidal stripping, which is related to the galaxy physical interactions (nurture). In such case, the compact object should reveal the properties of the central region of the progenitor, which would be a larger and more massive type of galaxy. Under this scenario, UCDs are thought to be the result of stripping dwarf galaxies (dEs) whereas cEs would be the stripped cores of larger ellipticals \citep{Faber1973} or disk galaxies (e.g. \citealt{Bekki2001}; \citealt{Graham2002}; \citealt{Drinkwater2003}). The other scenario represents an intrinsic process, where the galaxy was formed as it is, with no stripping involved, and thus is related to the nature of the system itself. Low mass UCDs would thus be the high-mass end of the globular cluster family (e.g. \citealt{Murray2009}; \citealt{Bruns2011}; \citealt{Mieske2012}), while cEs would be the unstripped low mass and faint end of ETGs (e.g. \citealt{Wirth1984}; \citealt{Nieto1987}; \citealt{Kormendy2009}; \citealt{Kormendy2012}; \citealt{Paudel2014b}; \citealt{Martinovic2017}). We emphasize here that when we refer to cEs being such low-mass ETGs, we mean compact elliptical galaxies with stellar masses of 10$^{8} \la$ M$_{*}$/M$_{\odot}\la$10$^{10}$ and high densities, and not dEs. The latter tend to have much lower densities, slightly larger sizes and smaller velocity dispersions. They represent a challenge by themselves, with their origins being still under strong debate (e.g. \citealt{Wirth1984}; \citealt{Kormendy1985}; \citealt{Bender1992}; \citealt{Graham2003}; \citealt{Ferrarese2006}; \citealt{Graham2017}; \citealt{Janz2017}).  

Although the tidal stripping origin seems to be a very common mechanism for shaping the population of compact objects, it is still unclear for which mass range and environment this dominates and what is the resulting abundance of cEs. Recently, \citet{Martinovic2017} analyzed a sample of compact dwarf galaxies in clusters from the cosmological simulations. Their results showed that they recover the two main mechanisms discussed in this paper (intrinsic \text{vs} stripped). However they found that the majority were formed in situ within the cluster. This is, they were already created as low mass compact objects and the cluster environment itself prevented further evolution (e.g. \citealt{Wellons2016}). Only 30\% of their sample were Milky-way type galaxies that formed outside the cluster environment and later fell into it, suffering tidal interactions that stripped their stars as they sank towards the center of the cluster. Whether this stands for less dense environments such groups and fields, is still to be determined.

\begin{figure*}
\centering
\includegraphics[scale=0.7]{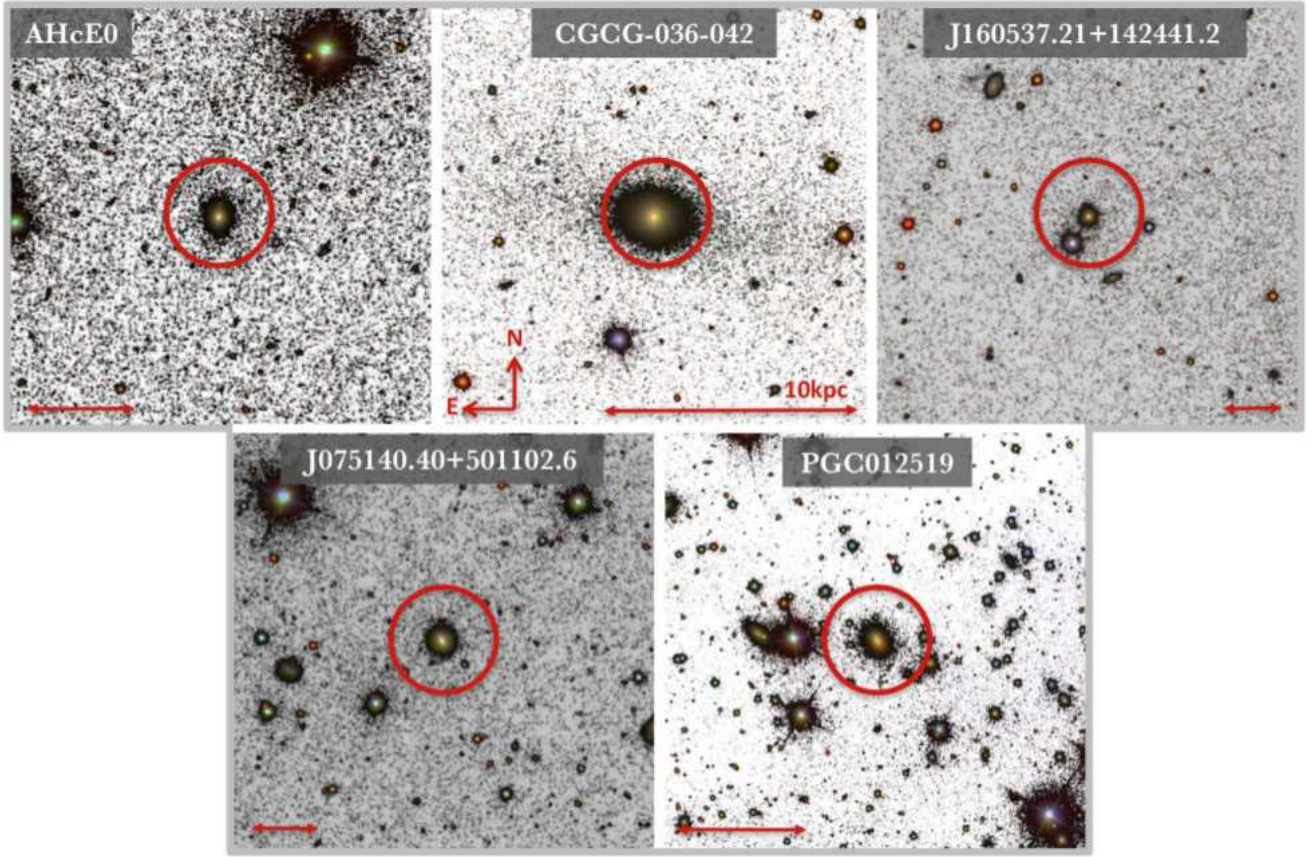}
\label{figure:1}
\caption
{SDSS stamps for the \textit{no host} galaxies, marked with the red circles. They show the low flux regions in greyscale and high flux regions in the original SDSS colour image, with a physical scale of 10\,kpc shown by the red bar. Note that the last two objects are located in the outskirts of clusters.}
\end{figure*}

\begin{figure*}
\centering
\includegraphics[scale=0.7]{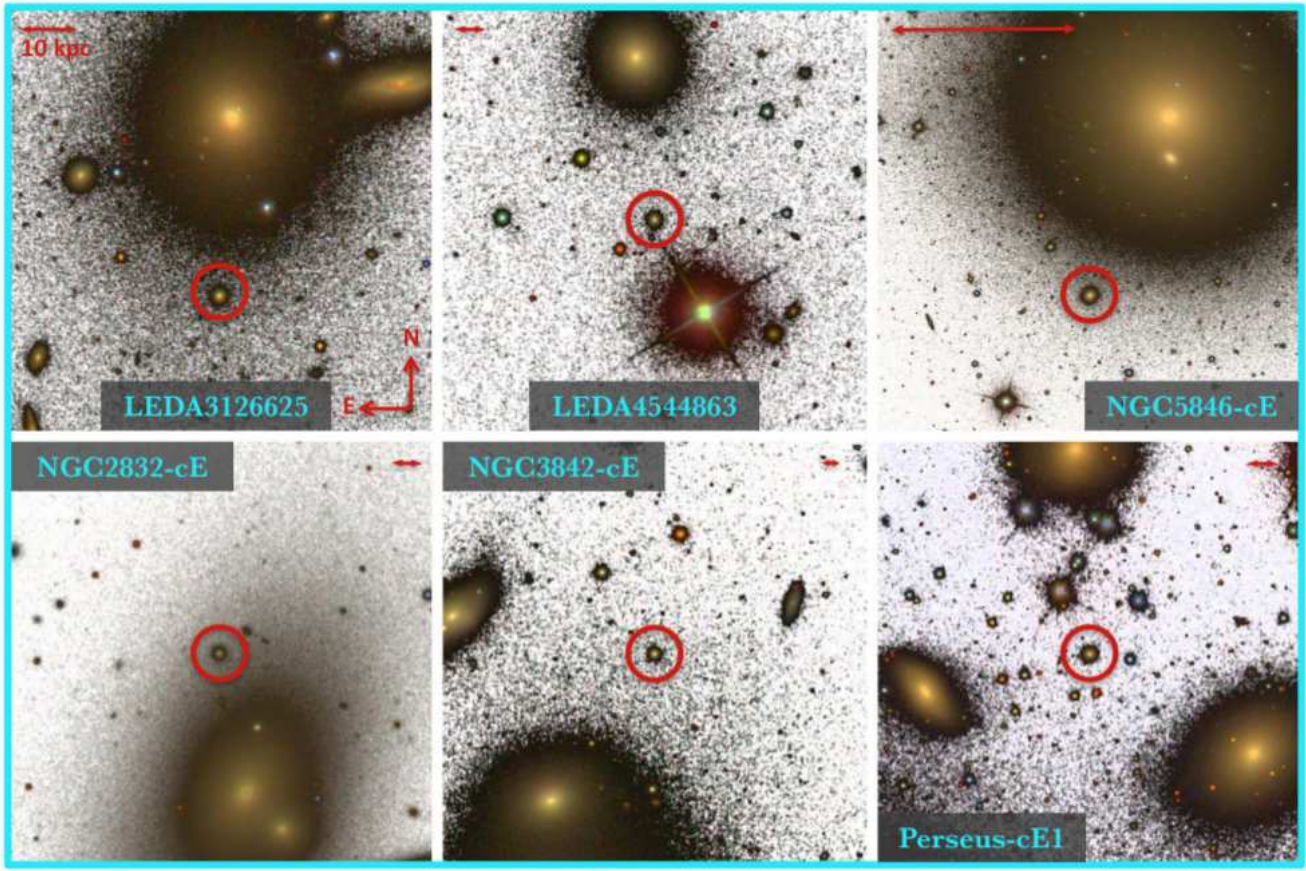}
\label{figure:2}
\caption
{SDSS stamps for the \textit{near host} galaxies. They appear to be undisturbed by the host, although their relative velocities confirm their association.As in Figure 1, the candidates are highlighted with a red circle and the physical scale of 10\,kpc is indicated by the red bar in each panel.}
\end{figure*}

\begin{figure*}
\centering
\includegraphics[scale=0.7]{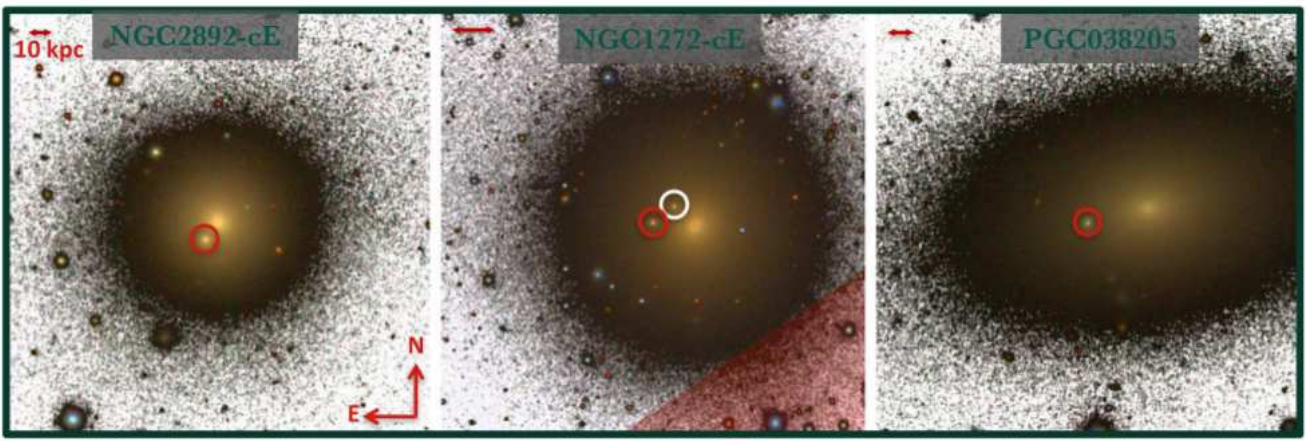}
\label{figure:3}
\caption
{SDSS stamps for the \textit{within halo} galaxies. They appear to be within the halo of the host galaxy, with projected distances of less than 20\,kpc  NGC\,1272 has two associated cEs, but we only analyse NGC1272-cE1 (red) in this work, as no spectroscopic data are available for NGC\,1272-cE2 (white circle). As in Figure 1, the candidates are highlighted with a red circle and the physical scale of 10\,kpc is indicated by the red bar in each panel.}
\end{figure*}

\begin{figure}
\centering
\includegraphics[scale=0.7]{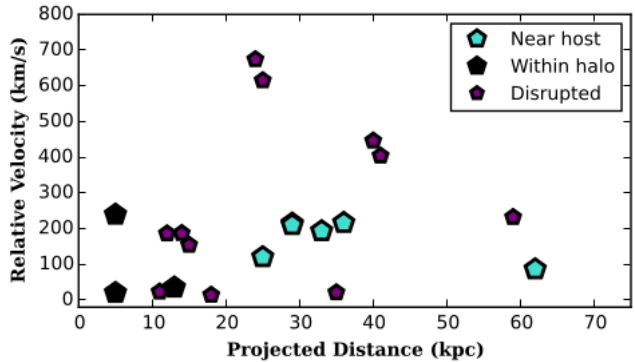}\\
\caption{Projected distance to the plausible host. \textit{Disrupted} galaxies (purple) are scattered all over the parameter space, but those classified as \textit{within halo} (black) are at a distance of less than 15\,kpc from the host, while those \textit{near host} (cyan) are at larger distances. For both classes, no signs of interaction are visually distinctive.}
\label{figure:4}
\end{figure}

\begin{figure*}
\centering
\includegraphics[scale=0.9]{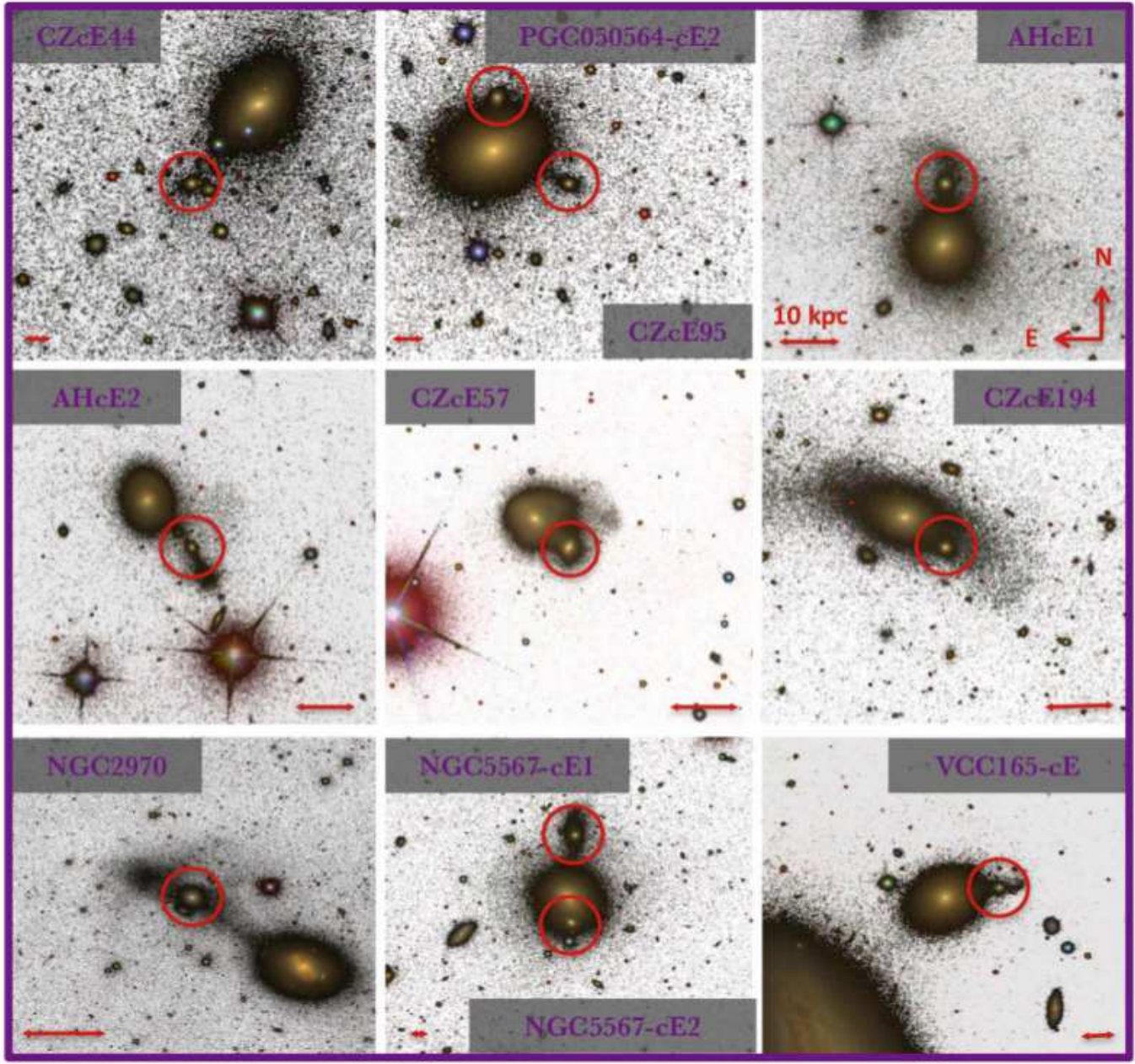}
\label{figure:5}
\caption
{SDSS stamps for the \textit{disrupted} galaxies. They are either embedded in streams or highly interacting with the host. In some cases, one host can have more than one associated cE. As in Figure 1, the candidates are highlighted with a red circle and the physical scale of 10\,kpc is indicated by the red bar in each panel.}
\end{figure*}

\begin{figure}
\centering
\includegraphics[scale=0.4]{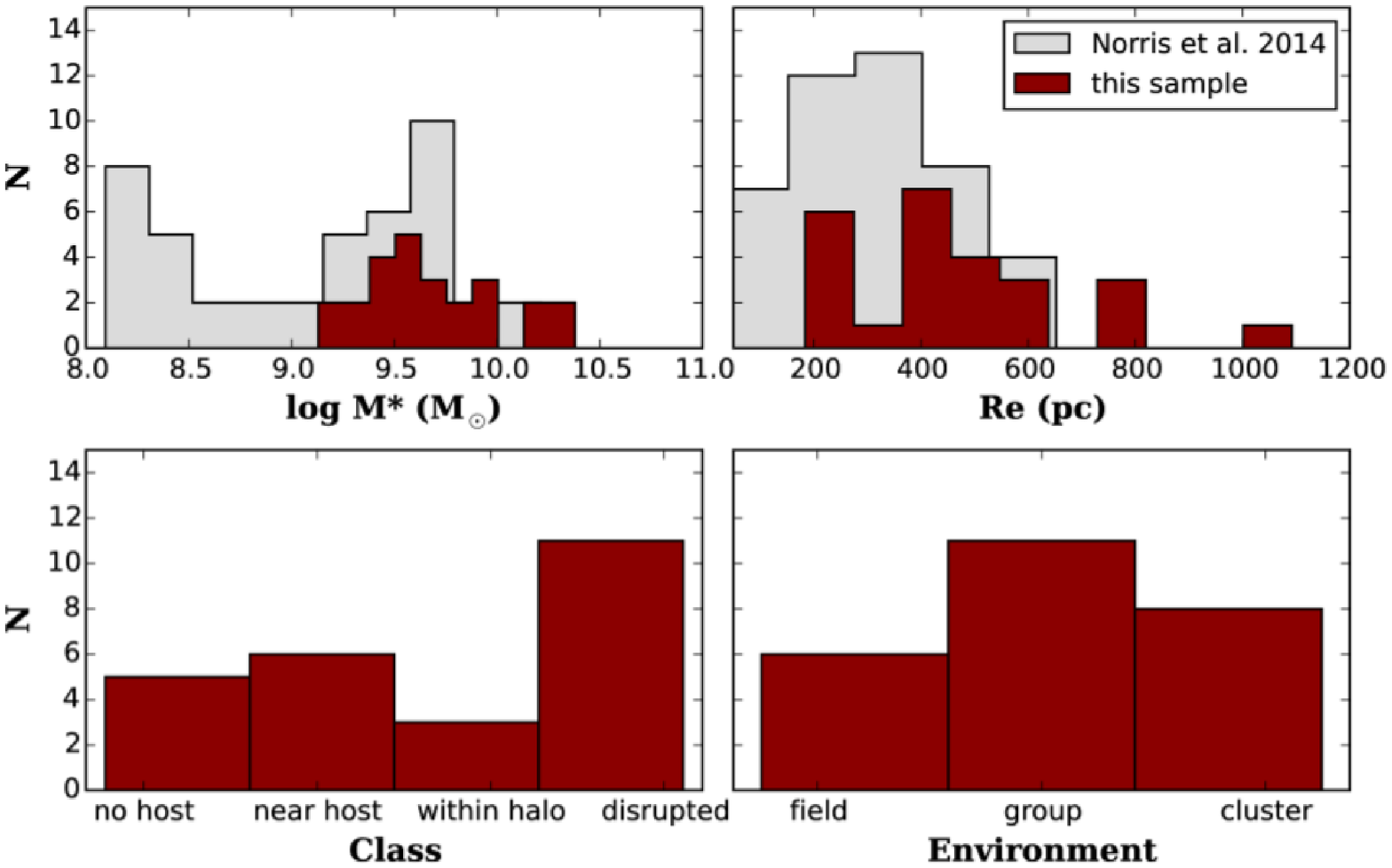}\\
\caption{Parameter coverage for the sample of 25 compact ellipticals studied in this paper. Upper panels show the stellar mass and size ranges covered by our sample and the one of N+14. We see that our sample covers the upper mass end of cEs, with a coverage of sizes more similar to the N+14 sample. Lower panels show the number counts of cEs in this sample for each class (left) and environment (right)}
\label{figure:6}
\end{figure}

Revealing such origins is not a trivial endeavour. Even for M32, the prototype of cE for excellence, its origin is not secure yet. Some works claim it as the ending product of the threshing of a larger spiral (e.g. \citealt{Bekki2001}), supported by the evidence for a faint, stripped remnant disk around it \citep{Graham2002}. However, other evidences seem to discard such stripping origin, provided by the analysis of the stellar populations of both M32 and M31 (the proposed host galaxy for M32; e.g. \citealt{Choi2002}). Furthermore, there are many other problems regarding the determination of the origins of cEs. It is almost impossible to determine in what exact evolutionary stage a galaxy is with simply visual tools. For example, if the cE is near a massive host but there are no signs of interaction, how can we be assured it has been already stripped of if the threshing is about to begin? Or when a galaxy is embedded in a halo or stream, can we assure it is being stripped or is it only a projection effect? Such a challenge can only be tackled by combining the information from different tools that allow one to discriminate between the possible origins (intrinsic \textit{vs} stripped). 

So far, most studies have focused on either one single property or they have been limited to a couple of galaxies per study. For example, the integrated stellar populations of compact stellar systems spanning a range of stellar masses were investigated by \citet{Janz2016}. They found that cEs tend to deviate from the mass-metallicity relation, being more metal rich than expected from their mass (i.e. with metallicities more typical to the cores of ETGs). A wide range of stellar ages and $\alpha$-enhancements was also reported for cEs, which would reflect the different evolutionary stages the progenitor galaxy experiences while being stripped. However, much further insight can be obtained by probing their star formation histories and formation timescales, which has not yet been tackled to date. 

Another interesting property that was found for compact stellar systems was that they show elevated dynamical to stellar mass ratios \citep{Forbes2014}. This behavior is stronger in low-mass cEs whereas the ratios are close to unity for the highest mass ones. The cause for such elevated ratios is still a subject of debate, as they could be due to several effects. One is the presence of a black hole in the galaxy, giving higher dynamical masses than expected (e.g. \citealt{Kormendy1997}; \citealt{Mieske2013}; \citealt{Seth2014}) or the presence of dark matter (e.g. \citealt{Hasegan2007}; \citealt{Baumgardt2008}). A different initial mass function (IMF) would also have a strong impact on the stellar mass derived (e.g.  \citealt{Ferre-Mateu2013}). However, the tidal stripping scenario tested here can also account for such behavior. While the more massive progenitor loses the majority of its stellar mass and shrinks, its velocity dispersion remains almost unaltered. Therefore, deviations from a M$_{\mathrm{dyn}}$/M$_{*}$ of unity can represent these evolving stages, and thus can also be used to discriminate between the possible origins of compact objects (e.g. \citealt{Pfeffer2013}, \citealt{Forbes2014}, \citealt{Janz2016b}).

In this paper we study the stellar populations, star formation histories and mass ratios of a sample of cEs for which good SDSS spectroscopy and photometric data are available. They have been visually classified to represent different evolutionary stages a galaxy can undergo under both possible origins (tidal stripping \textit{vs} intrinsic). However, such visual classification alone cannot securely describe the stage of each galaxy, as projection effects and other caveats might mask the true stage of the cE. In order to overcome such caveats, we use a set of discriminant tools which, if combined, can help constraining the formation mechanism, evolutionary stage and progenitor type for the cEs in our sample. Dependence on the environment will also be addressed, as our sample comprises galaxies in the field, in groups and in galaxy clusters. 
Throughout this work we adopt a standard cosmological model with the following parameters: $H_0$=69 km s$^{-1}$ Mpc$^{-1}$, $\Omega_m$=0.3 and $\Omega_\Lambda$=0.7.

\section{Sample}
We aim to create a sample representative of the different evolutionary stages in the formation process of cEs. We thus look in the SDSS for literature cEs with available spectroscopy. Because high S/N is crucial for deriving robust star formation histories (SFHs) and stellar populations, we impose a minimum S/N of 20 per \AA  \, (e.g. \citealt{CidFernandes2014}; \citealt{GonzalezDelgado2014}; \citealt{Ferre-Mateu2014}) in the SDSS spectral region corresponding to the $r$-band. 

We are mostly interested in the galaxies that are 'caught in the act', showing signs of tidal stripping, as they can provide direct clues about the different formation channels. Therefore, we first search for such disrupted candidates in the sample of cEs from \citet{Chilingarian2015} (CZ+15 hereafter). Only eight out of their $sim$200 objects were reported to be currently in a stripping process, but three of them do not have enough S/N and thus have been excluded. Two others are near the limit of the S/N criterion in the $r$-band (CZcE44 and CZcE95) but have enough signal in the other spectral regions. We have included them in our sample but flagged them as questionable for the stellar populations analysis. The remaining three objects have higher S/N and are therefore considered here (CZcE57, CZcE194, CZcE181). The latter was also studied by \citet{Huxor2011} as AHcE1, together with another cE showing tidal features, both with good quality SDSS spectra (AHcE2). \citet{Huxor2013} found another cE in the SDSS, AHcE0. This is a remarkable case since it is isolated and therefore an interesting case for our sample. While such isolated cEs are rare, other candidates have been previously studied, such as CGCG-036-042 \citep{Paudel2014b}, also included here.

We search for other good cE candidates from the AIMSS project (\citealt{Norris2014};  N+14 hereafter). This is a compilation of compact stellar systems spanning a large range of mass and sizes, hence another excellent source for candidates. We find 10 more objects (out of their $\sim$50 galaxies defined as cEs), for which SDSS spectra with S/N $\ge$ 20 exist. These are LEDA\,3126628, LEDA\,4544863, J075140.40+501102.6 (J07+50 hereafter), J160537.21+142441.2 (J16+14 hereafter), NGC\,1272-cE1, NGC\,2832-cE, NGC\,2892-cE, NGC\,5846-cE, PGC\,038205 and VCC165-cE.

In addition, seven new objects are reported here as cEs for the first time, found through SQL searches. Automated searches for cEs in SDSS spectroscopic catalogs were first pioneered by \citet{Chilingarian2009} and CZ+15, and we are developing new search techniques along similar lines (Dixon et al., in prep.). We thus use parameters previously calibrated by known cEs, imposing a redshift range (0.007$< z <$0.035), a stellar mass range (9.0 $< logM <$\,13.0) and most importantly, a very restrictive size range (0.02 $< \mathrm{R_e (kpc)} <$ 0.6). For the sizes we use a weighted average of the de Vaucouleurs and exponential model results from SDSS, as described in Section 3.1. With these parameters, roughly half of the $\sim$\,150 objects selected could be rejected by eye as corresponding to dense nuclei in large galaxies. The rest appeared visually as potential cEs, which would require more careful analysis to be confirmed. From them, only four of them had enough S/N to be included in the sample: PGC\,012519, NGC\,3842-cE, Perseus-cE1, NGC\,5567-cE1. Three others (PGC\,050564-cE2, NGC\,5567-cE2 and NGC1272-cE2) were noticed visually as additional companions to the host galaxies of previously discovered cEs (CZcE95, NGC\,5567-cE1 and NGC\,1272-cE1, respectively), but we include only PGC\,050564-cE2 and NGC\,5567-cE2 in our sample, due to the lack of spectra for NGC\,1272-cE2.

Finally, we include one more object that also has sufficient S/N SDSS spectrum: NGC\,2970. Despite having been classified as a dE \citep{Paudel2014}, its reported size and stellar mass are very close to our limiting criteria. What is most interesting is that it is located in a stream, showing signs of being currently stripped. What it is unknown for now is whether this stripping is in its late stages (and thus NGC\,2970 will remain as a dE), or whether this is only an early stage and further evolution towards the cEs or UCDs regimes is expected.

This gives a sample of 25 cE galaxies with SDSS spectra with S/N $\ge$20. They have been visually classified from the SDSS images into four classes that represent different evolutionary stages in the formation processes considered here. The first class is the \textit{no host} cEs (Figure 1), which are galaxies that cannot be associated to any nearby galaxy (no plausible host within a radius of several hundred kpc). They can represent either a stripped system that was ejected from a denser association or they can be the unstripped low-mass end of ETGs, thus having an intrinsic origin. There are also objects nearby a larger galaxy that is a plausible host from their relative velocities, but there are no visual signs of interaction between them. In such cases, the visual inspection alone cannot determine their origin. They could represent either the latest stages of the stripping process, where the progenitor has been completely stripped and only the remnant is left, or they could also represent an early stage, as a compact system that is about to become disrupted and for which further evolution towards a more compact and less massive system is expected. Such galaxies can be subdivided into two different classes. There are those that are outside the influence of the host but clearly associated to it due to their relative velocities (\textit{near host}; Figure 2) and those that lie within the halo of the host but that show no distinguishable signs of interaction (\textit{within halo}; Figure 3). A further check in deeper images (e.g. PAN-STARRS1) when available, was performed to find any low surface tidal features that could be lost on the shallower SDSS images. However, nothing was found for any of the galaxies in these two last classes. Figure 4 shows the projected distance between the cE and its host with respect to their relative velocities, as quoted in Table 1. It is seen that those we selected as \textit{within halo} are at distances of $\la$15\,kpc from their hosts, while those \textit{near host} are further away. However, we see that the \textit{disrupted} class is found at all projected distances. The last class represents those objects that are caught in the act (\textit{disrupted}; Figure 5). These can be highly interacting with their host or embedded in streams, but in all cases, the tidal stripping scenario is clear as it is currently happening. 

The 25 objects also cover a variety of environments: field, groups or clusters. They have been classified by their {\tt SIMBAD} environment, in a similar fashion to N+14. We consider \textit{field} galaxies if they are isolated or are in associations of fewer than 5 galaxies. \textit{Groups} are defined as galaxies in associations with more than 5 galaxies and fewer than 50, and galaxies in \textit{clusters} are associations that contain more than 50 galaxies. It is important to highlight that isolated or field galaxies do not mean galaxies with \textit{no host}, but merely represent the lowest density environments. In addition, this global environment is not representative of the local one in some cases. For example, galaxies in the outskirts of clusters have an environment that is more similar to the field one. We will come back to these particular cases later.

Figure 6 presents the parameter coverage of the sample. Upper panels show the stellar masses and sizes (from Section 3) compared to the cEs in the AIMSS sample (N+14). Our sample represents about half of N+14 and shows a particularly good coverage at the massive end, where it  is more difficult to determine whether the objects have an intrinsic or stripped origin and their evolutionary stage. The lower panels show the coverage of the four galaxy classes and environments considered in this work. Most of the galaxies belong to the \textit{disrupted} class, which is a natural result of how they were selected. They are the ones we are most interested in, as their origin is clear and thus we can use their properties to compare with the other classes with unknown origins. 

\begin{turnpage}
\begin{table*}
\centering
\label{table:1}                      
\caption{Compact Ellipticals sample}
\setlength{\tabcolsep}{3pt}   
\begin{tabular}{c | c c c c c c c c c c}   
\toprule      
  \textbf{Galaxy}  &   \textbf{R.A.} & \textbf{Dec.} & \textbf{$D$ ($z$)} &  \textbf{S/N} & \textbf{host} & \textbf{Proj. D} & \textbf{$\Delta$vr} & \textbf{class} & \textbf{environ} & \textbf{source} \\  
                            &    (J2000)       &   (J2000)      & (Mpc)                    &                     &                    &      (kpc)                       &    (km\,s$^{-1}$)             &                        &                          &                          \\
\midrule
\midrule
AHcE0                & 09:47:29.2 &  14:12:45.3 &  80.6 (0.019) &  33 & -             & -   &  -    & \textit{no host}     & Field   &   H+13/J+16        \\
CGCG-036-042         & 10:08:10.3 &  02:27:48.2 &  80.6 (0.019) &  54 & -             & -   &  -    & \textit{no host}     & Field   &   J+16             \\
J160537.21+142441.2  & 16:05:37.2 &  14:24:41.3 &  68.1 (0.016) &  43 & -             & -   &  -    & \textit{no host}     & Field   &   J+16             \\ 
J075140.40+501102.6  & 07:51:40.4 &  50:11:02.6 &  84.8 (0.020) &  28 & -             & -   &  -    & \textit{no host}     & Cluster &   J+16             \\
PGC012519            & 03:20:32.9 &  41:34:26.8 &  42.9 (0.010) &  87 & -             & -   &  -    & \textit{no host}     & Cluster &   SDSS             \\
\midrule                                                                                                    
LEDA\,3126625        & 01:49:14.4 &  13:01:55.0 &  68.1 (0.016) &  45 & NGC677        & 29  &  214  & \textit{near host}   & Group   &   J+16             \\
LEDA\,4544863        & 13:38:42.4 &  31:14:57.1 &  63.9 (0.015) &  26 & MCG+05-32-049 & 25  &  120  & \textit{near host}   & Group   &   J+16             \\
NGC5846-cE           & 15:06:34.2 &  01:33:31.7 &  21.5 (0.005) &  56 & NGC5846       & 36  &  216  & \textit{near host}   & Group   &   J+16             \\
NGC2832-cE           & 09:19:47.8 &  33:46:04.8 &  97.1 (0.023) &  39 & NGC2832       & 33  &  193  & \textit{near host}   & Cluster &   J+16             \\
NGC3842-cE           & 11:43:58.7 &  19:59:28.2 &  88.9 (0.021) &  29 & NGC3842       & 62  &   86  & \textit{near host}   & Cluster &   SDSS             \\
Perseus-cE1          & 03:19:33.6 &  41:33:12.8 &  72.3 (0.017) &  35 & NGC1273       & 26  &  210  & \textit{near host}   & Cluster &   SDSS             \\
\midrule                                                                                    
NGC2892-cE           & 09:32:53.9 &  67:36:54.6 &  93.0 (0.022) &  23 & NGC2892       &  5  &   20  & \textit{within halo} & Group   &   J+16             \\
NGC1272-cE1          & 03:19:23.1 &  41:29:28.1 &  55.6 (0.013) &  41 & NGC1272       &  5  &  239  & \textit{within halo} & Cluster &   J+16             \\
PGC038205            & 12:04:28.9 &  01:53:38.8 &  88.9 (0.021) &  52 & NGC4073       & 13  &   35  & \textit{within halo} & Cluster &   J+16             \\
\midrule                                                                                    
CZcE44               & 10:05:15.6 &  50:10:14.6 & 204.5 (0.050) &  18 & LEDA\,2365336 & 59  &  232  & \textit{disrupted}   & Field   &   CZ+15            \\
CZcE95               & 14:09:56.8 &  54:52:35.4 & 177.4 (0.043) &  19 & PGC050564     & 40  &  446  & \textit{disrupted}   & Field   &   CZ+15            \\
PGC050564-cE2        & 14:10:01.3 &  54:53:24.1 & 165.6 (0.040) &  25 & PGC050564     & 24  &  674  & \textit{disrupted}   & Field   &   SDSS             \\
AHcE1/CZcE181        & 11:04:04.4 &  45:16:18.9 &  88.9 (0.021) &  29 & PGC033435     & 15  &  154  & \textit{disrupted}   & Group   &   H+11/J+16/CZ+15  \\
AHcE2                & 23:15:12.6 & -01:14:58.3 & 105.4 (0.025) &  21 & III Zw097     & 29  &   14  & \textit{disrupted}   & Group   &   H+11/J+16        \\
CZcE57               & 07:59:05.1 &  27:27:34.3 &  93.0 (0.022) &  40 & IC2213        & 11  &   23  & \textit{disrupted}   & Group   &   CZ+15            \\
CZcE194              & 12:10:31.1 &  00:40:21.9 &  80.6 (0.019) &  28 & PGC038740     & 12  &  186  & \textit{disrupted}   & Group   &   CZ+15            \\
NGC2970              & 09:43:31.1 &  31:58:37.1 &  21.5 (0.005) &  40 & NGC2698       & 41  &  404  & \textit{disrupted}   & Group   &   SDSS             \\
NGC5567-cE1          & 14:19:17.2 &  35:09:15.2 & 113.5 (0.027) &  21 & NGC5567       & 14  &  187  & \textit{disrupted}   & Group   &   SDSS             \\
NGC5567-cE2          & 14:19:17.4 &  35:07:53.4 & 117.6 (0.028) &  38 & NGC5567       & 35  &   21  & \textit{disrupted}   & Group   &   SDSS             \\
VCC165-cE            & 12:15:51.2 &  13:13:03.4 & 173.5 (0.042) &  20 & VCC165        & 25  &  615  & \textit{disrupted}   & Cluster &   J+16             \\
\bottomrule
\end{tabular}

\vspace{0.2cm}
{Sample of compact ellipticals for which high quality spectroscopy (S/N\,>20) in the SDSS is available. The source for the selection parameters is quoted as H+11 for \citet{Huxor2011}, H+13 for \citet{Huxor2013}, CZ+15 for \citet{Chilingarian2015} and J+16 for \citet{Janz2016}. For the latter, they were initially selected from the AIMSS project of \citet{Norris2014}. Distances are derived from the galaxy's redshift with the assumed cosmology, with a double check from the host redshift. The classification into one of the four classes is separated in the table by the four blocks: with \textit{no host}, with host but unknown origin (\textit{near host}), those that lie within the halo of the host but disruption is unclear (\textit{within halo}), and those interacting with the host or embedded in streams (\textit{disrupted}). If the host is known, it is stated, together with the projected distance from the host to the cE and their relative velocities (as shown in Figure 4). The galaxy global environment is also presented, as described by the catalogs in SIMBAD. }
\end{table*}
\end{turnpage}

\section{ANALYSIS}
\subsection{Sizes, kinematics and dynamical masses}
\begin{table}
\centering
\label{table:2}                      
\caption{Physical properties and kinematics}    
\begin{tabular}{c | c c c c}   
\toprule
\textbf{Galaxy}   & \textbf{ $\mathrm{R_e}$} & \textbf{$\mathrm{R_{fibre}/R_e}$ }   & \textbf{$\sigma$} & \textbf{M$_{\mathrm{dyn}}$ } \\  
                  & (pc) &  & (km\,s$^{-1}$)      &($\times$10$^{9}$M$_{\odot}$)   \\
\midrule	
\midrule			 
AHcE0            &  499       & 1.2 &  112.9$\pm$\,2.3  & 9.6  \\
CGCG-036-042     &  465       & 0.4 &  106.0$\pm$\,1.7  & 7.8  \\
J16+14           &  511       & 1.0 &   61.4$\pm$\,1.6  & 2.9  \\
J07+50           &  185       & 3.5 &  123.6$\pm$\,3.0  & 4.2  \\
PGC012519        &  398       & 0.8 &  222.2$\pm$\,2.3  & 29.7 \\
\midrule                     
LEDA\,3126625    &  414       & 1.2 &   62.7$\pm$\,1.9  & 2.4  \\
LEDA\,4544863    &  433       & 1.1 &   58.9$\pm$\,4.9  & 3.1  \\
NGC5846-cE       &  240       & 0.7 &  114.4$\pm$\,2.9  & 7.2  \\ 
NGC2832-cE       &  375       & 1.9 &  114.4$\pm$\,2.6  & 7.4  \\
NGC3842-cE       &  184(\dag) & 3.6 &   78.2$\pm$\,2.9  & 1.7  \\
Perseus-cE1      &  452(\dag) & 1.2 &   71.8$\pm$\,2.2  & 3.5  \\
\midrule   
NGC2892-cE       &  580       & 1.2 &  137.7$\pm$\,2.4  & 16.6 \\
NGC1272-cE1      &  377       & 1.0 &   82.4$\pm$\,3.3  & 3.8  \\ 
PGC038205        &  616       & 1.1 &  182.2$\pm$\,4.1  & 30.9 \\
\midrule   
CZcE44           &  199(\dag) & 3.1  &  106.9$\pm$\,4.6  & 8.5  \\
CZcE95           &  524(\dag) & 2.6  &   90.5$\pm$\,3.1  & 6.4  \\
PGC050564-cE2    &  775(\dag) & 1.6  &  102.4$\pm$\,2.8  & 12.3 \\
AHcE1            &  338       & 2.0  &   91.8$\pm$\,2.2  & 4.3  \\
AHcE2            &  263       & 3.0  &  108.5$\pm$\,5.0  & 4.6  \\
CZcE57           &  775(\dag) & 0.9  &  169.3$\pm$\,3.3  & 33.7 \\
CZcE194          &  491(\dag) & 1.3  &   72.1$\pm$\,2.1  & 3.8  \\ 
NGC2970          &  793(\dag) & 0.2  &   47.7$\pm$\,1.6  & 2.7  \\
NGC5567-cE1      &  367(\dag) & 2.4  &   65.7$\pm$\,3.5  & 2.4  \\
NGC5567-cE2      & 1092(\dag) & 0.8  &  141.8$\pm$\,2.0  & 33.2 \\ 
VCC165-cE        &  200       & 6.6  &   88.8$\pm$\,1.9  & 2.3  \\
\bottomrule	                            
\end{tabular}

\vspace{0.2cm}
{Physical properties of the sample. The different parts of the table correspond to the four visual classes as in Table 1 (\textit{no host, near host, within halo} and \textit{disrupted}). Sizes represent the published values unless marked with \dag, in which case the SDSS value in the $r$-band is used. The radial coverage of the SDSS fibre corresponding to a $\mathrm{R_{fibre}}$=\,1.5" is also shown. The quoted velocity dispersions are those obtained with {\tt pPXF}, used to derive the quoted dynamical mass.}
\end{table}

\begin{figure}
\centering
\includegraphics[scale=0.48]{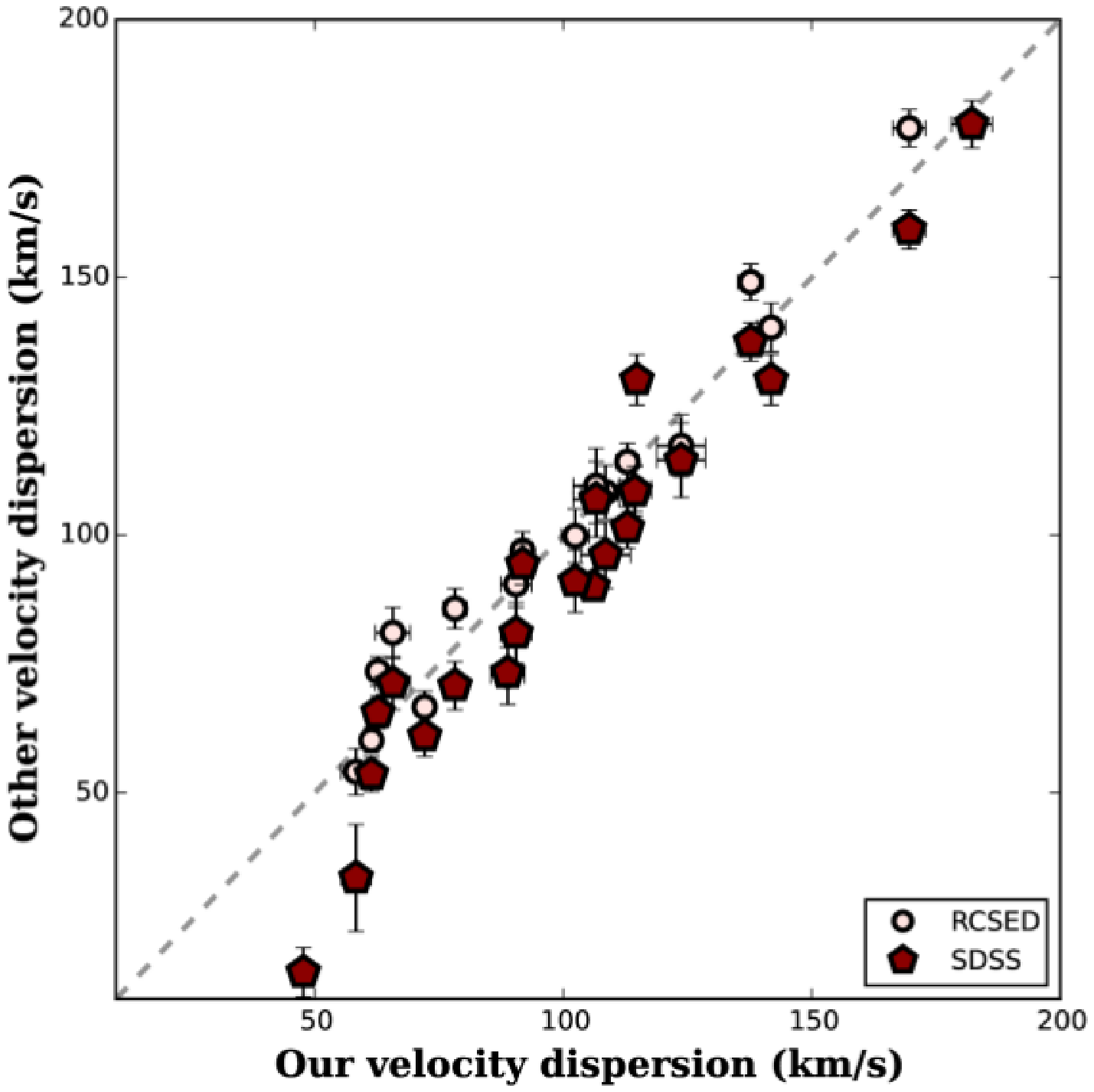}\\
\caption{Comparison of the velocity dispersions from different sources or methods. We compare those used in this paper, which have been derived with {\tt pPXF}, with the ones catalogued by SDSS (red pentagons) and those from RCSED (\citealt{Chilingarian2017}; open circles). The grey dashed line is the 1:1 relation to guide the eye.}
\label{figure:7}
\end{figure}

We use the published galaxy sizes, if available, which have been measured by modelling higher resolution images than those provided by SDSS (see Table 2). But for those cEs with no literature available we use their SDSS values and proceed cautiously as such sizes can be overestimated. Following the SDSS guidelines (\textit{http://www.sdss.org/dr12/algorithms/magnitudes/}), we consider the values derived from the {\tt modelMag}. It is the best of the two fits between a de Vaucouleurs ({\tt deVRad}) model and an exponential model ({\tt expRad}), both in the $r$-band. These models produce quantities such as the total magnitudes, effective radii (the one we consider here), axis ratios and position angles. However, we shown in Appendix A that this method can overestimate such quantities, in particular for those cEs that are embedded in their host galaxy or that are at further distances (i.e. $z\,\ga$ 0.025, see e.g. \citealt{Huxor2011}). In such cases we consider the size as an upper limit and the galaxies have been treated separately when studying the different discriminant tools.

Another caveat to consider is the restricted coverage of the 3" SDSS fiber, as presented in Table 2. Because our candidates cover a range of redshifts and sizes, the fraction of light inside the fiber will also vary, which can have an effect on the derived stellar populations (e.g. \citealt{Poggianti2004}; \citealt{McDermid2015}). While the redshift range covered does not introduce relevant systematic effects (e.g. \citealt{Kewley2005}), the different coverages due to different galaxy sizes need to be accounted for. Using the {\tt ATLAS3D} sample, it has been seen that typically larger coverages tend to provide lower metallicities (e.g. $\sim$0.2\,dex). However, ages tend to be more robust, with those with consistently old ages ($\ga$10\,Gyr) showing almost no variation, although the stellar populations tend to get older with larger apertures for integrated values of $\la$5\,Gyr (the younger contributions are concentrated in the centers; \citealt{McDermid2015}). This will be taken into proper account in the Discussion Section.

We have three different estimates for the stellar kinematics of the galaxies. There is the velocity dispersion provided by SDSS and the one provided from RCSED (\citealt{specphot_rcsed16}; \citealt{Chilingarian2017}). The latter is a service that provides spectra and photometry of galaxies derived from cross-matches between SDSS, \textit{GALEX}, and UKIDSS catalogs. We have also measured the velocity dispersion with {\tt pPXF} \citep{Cappellari2004}. Figure 7 compares our measurements with those from the literature, showing a good agreement between the SDSS and the {\tt pPXF} ones, although the SDSS ones appear to be systematically lower. We remind the reader that those below $\sim$70\,km\,s$^{-1}$ are below the SDSS instrumental resolution. We will use the newly measured {\tt pPXF} values hereafter, which have been computed in the 3800 to 7400\,\AA\, spectral range using the MILES library of stellar spectra as templates \citep{Sanchez-Blazquez2006}. Using the rest of the SDSS spectral range (up to $\sim$9000\,\AA\,) only adds noise to the spectra, giving worse fits. This could explain the systematically higher values we obtain compared to SDSS. 

Having both the sizes and velocity dispersions of the systems, we can derive the dynamical masses under the assumption of pressure-dominated systems: \\
$M_{dyn}\,=\,C\,G^{-1}\,\sigma^{2}\, R$, \\
with $\sigma$ being the velocity dispersion, \textit{R} the size of the system and \textit{C} the virial coefficient. The latter value is basically determined by the S\'ersic index of the system \citep{Bertin2002}. For our galaxies, similar to UCDs and other compact systems, \textit{C}=\,6.5 is a fair approximation (\citealt{Mieske2013}; \citealt{Forbes2014}). Using this formula, we obtained new dynamical masses, reported in Table 2.     \begin{figure*}

\centering
\includegraphics[scale=0.42]{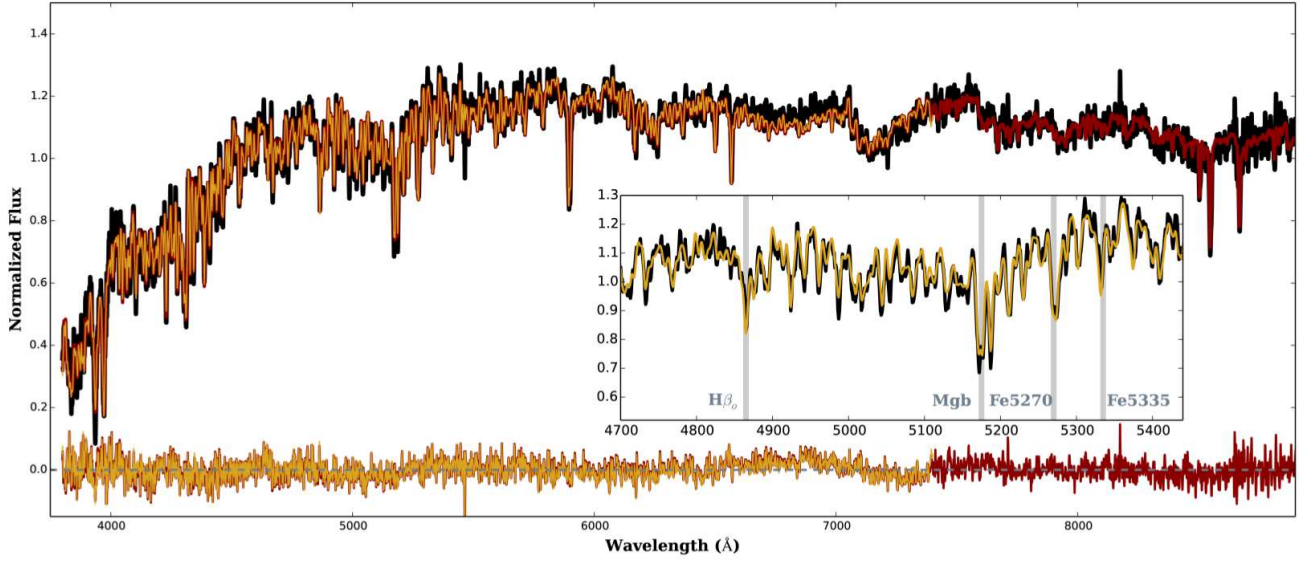}
\caption{Illustration of the methodology used to derive the stellar population parameters. The entire SDSS spectrum for one of our cEs is shown in black, with the fits from the {\tt STARLIGHT} full-spectral-fitting code overplotted. Red shows when fitting the entire spectra, while yellow represents the fitting for the spectral range used in the kinematical analysis, with their corresponding residuals around the grey line. The inset is a zoom-in of the spectral region of interest for the line indices approach, highlighting the main line indices used in Appendix B to obtain the [$\alpha$/Fe] ratios.}
\label{figure:8}
\end{figure*}

\begin{figure*}
\centering
\includegraphics[scale=1.1]{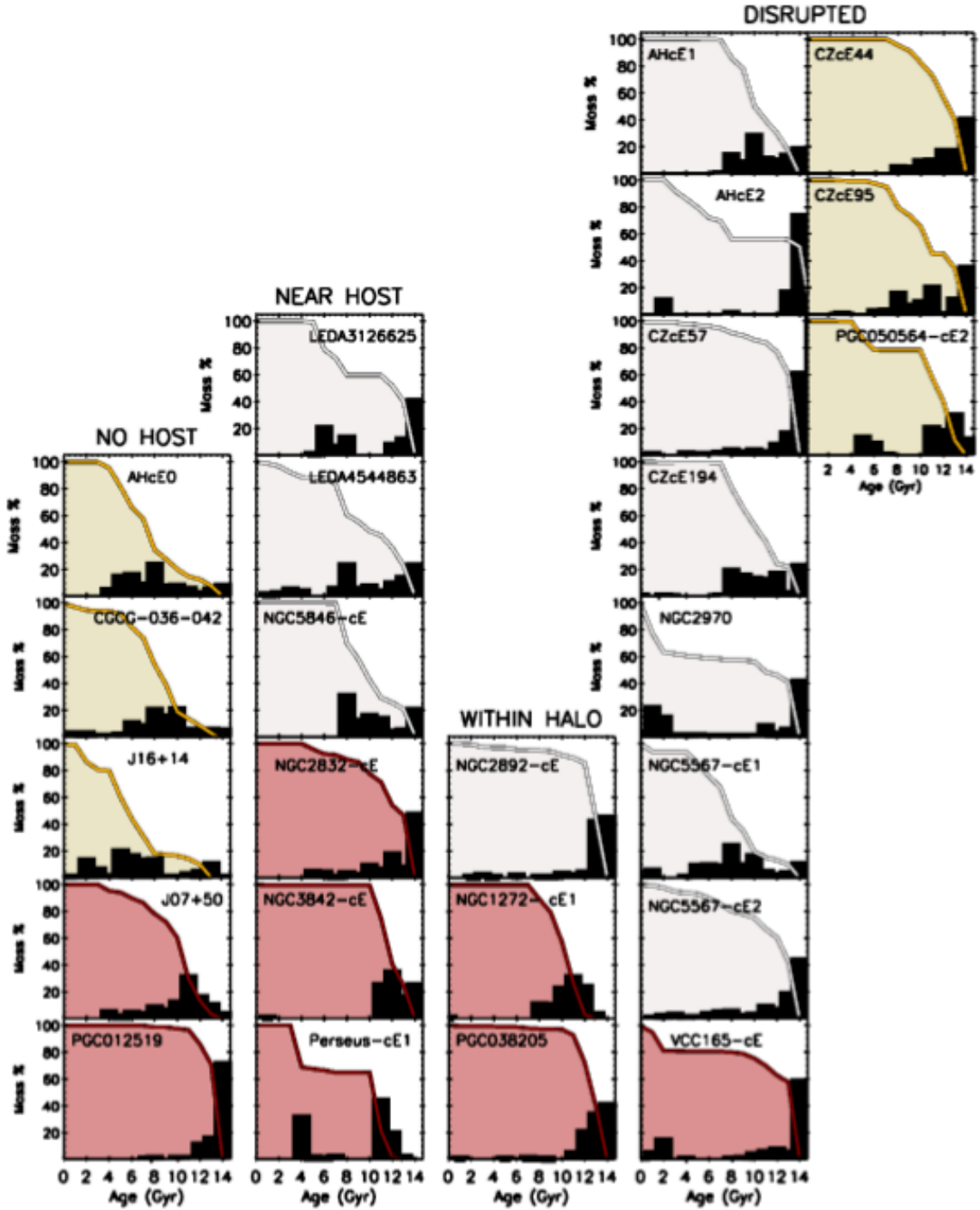}
\vspace{-0.2cm}
\caption{Star formation history  and cumulative stellar mass for each galaxy. The different classes are shown in different columns whereas galaxy environment has been colour-coded: yellow for field galaxies, red for group and grey for cluster galaxies. The cumulative stellar mass is a good indicator of how fast/slow the galaxy has built up its stellar mass. It can be seen that field galaxies typically show slower and more extended formation epochs that started later in time, while group and cluster galaxies show the earliest star forming events, albeit having a variety of SFHs.}
\label{figure:9}
\end{figure*}

\begin{figure}
\centering
\includegraphics[scale=0.52]{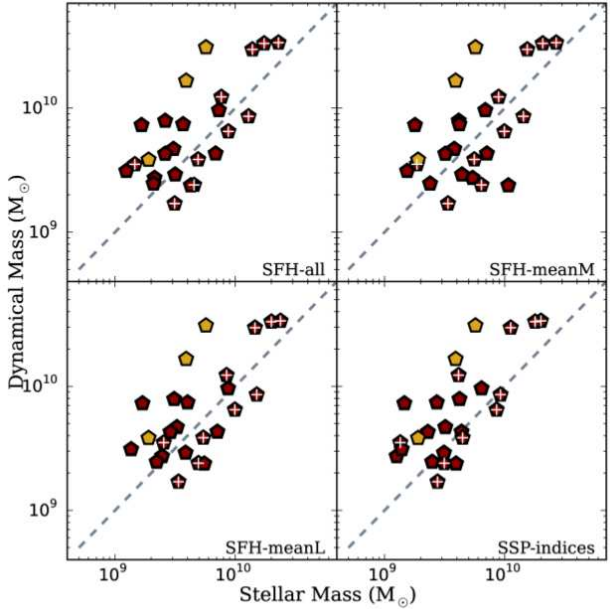}\\
\caption{Dynamical \textit{vs} stellar mass for the sample of cEs in this study. Each panel shows the stellar mass derived from different stellar M/L ratios, with a grey dashed line representing the 1:1 relation. Galaxies with a white cross are those for which the SDSS size was used to derive the dynamical mass, and the three yellow diamonds correspond to the three galaxies \textit{within halo} for which we consider their previously published stellar masses in J+16 due to high uncertainties in the SDSS model magnitudes.}
\label{figure:10}
\end{figure}
 
\begin{table*}\centering
\label{table:3}                      
\caption{Stellar population properties of cEs}    
\begin{tabular}{@{}c | ccc c ccc c cc@{}}\toprule 
  \textbf{Galaxy}  &\multicolumn{3}{c}{\textbf{Line indices}} & \phantom{abc}& \multicolumn{3}{c}{\textbf{Mass-weighted}} &\phantom{abc}&  \multicolumn{2}{c}{\textbf{Luminosity-weighted}}   \\
\cmidrule{2-4} \cmidrule{6-8} \cmidrule{10-11}
             &   Age &  [Z/H]    &   [$\alpha$/Fe]  &&  Age   & [Z/H]  & T50    &&  Age       & [Z/H] \\  
             &   (Gyr)& (dex)    &   (dex)               && (Gyr)  & (dex)   & (Gyr)  &&  (Gyr)    &  (dex) \\
\midrule		
\midrule	 
AHcE0             &  9.24$^{+0.46}_{-0.34}$ & $-$0.39$^{+0.02}_{-0.05}$ & $-$0.04$^{+0.01}_{-0.02}$ &&  9.7$\pm$\,2.6& $-$0.05$\pm$\,0.03 &  6.5 &&  9.3$\pm$\,2.5& $-$0.05$\pm$\,0.03   \\
CGCG-036-042      &  -                      &   -                       & $-$0.03(\ddag)            &&  8.7$\pm$\,1.5&    0.17$\pm$\,0.02 &  1.9 &&  6.2$\pm$\,1.0&    0.04$\pm$\,0.02   \\ 
J16+14            &  3.86$^{+2.22}_{-0.45}$ &    0.13$^{+0.10}_{-0.10}$ &    0.11(\ddag)            &&  6.6$\pm$\,1.5& $-$0.00$\pm$\,0.02 &  8.4 &&  5.3$\pm$\,1.2& $-$0.00$\pm$\,0.02   \\
J07+50            &  9.42$^{+3.58}_{-4.42}$ & $-$0.37$^{+0.07}_{-0.07}$ &    0.30$^{+0.16}_{-0.21}$ && 10.3$\pm$\,3.1& $-$0.04$\pm$\,0.04 &  4.0 &&  8.8$\pm$\,2.6&    0.04$\pm$\,0.04   \\
PGC012519         &  $>$ 15.00              & $-$0.23$^{+0.12}_{-0.10}$ &    0.25$^{+0.05}_{-0.05}$ && 13.5$\pm$\,0.5&    0.21$\pm$\,0.11 &  0.9 && 14.8$\pm$\,0.5&    0.19$\pm$\,0.11   \\
\midrule                                                                                                                                               
LEDA\,3126625     & 12.55$^{+7.45}_{-6.25}$ &    0.01$^{+0.08}_{-0.11}$ &    0.07$^{+0.14}_{-0.13}$ && 10.9$\pm$\,2.3&    0.17$\pm$\,0.00 &  1.9 && 10.2$\pm$\,2.2&    0.15$\pm$\,0.05   \\
LEDA\,4544863     &  8.64$^{+8.36}_{-5.14}$ &    0.10$^{+0.05}_{-0.08}$ & $-$0.02$^{+0.12}_{-0.17}$ && 10.1$\pm$\,3.1&    0.23$\pm$\,0.05 &  3.8 && 10.1$\pm$\,3.1& $-$0.08$\pm$\,0.05   \\
NGC5846-cE        &  $>$ 15.00              & $-$0.24$^{+0.06}_{-0.10}$ &    0.07(\ddag)            && 11.4$\pm$\,1.8&    0.20$\pm$\,0.03 &  4.1 && 11.6$\pm$\,1.8&    0.20$\pm$\,0.03   \\
NGC2832-cE        &  $>$ 15.00              & $-$0.25$^{+0.20}_{-0.25}$ &    0.23$^{+0.18}_{-0.17}$ && 12.1$\pm$\,2.9&    0.09$\pm$\,0.02 &  1.7 && 10.4$\pm$\,2.5&    0.17$\pm$\,0.02   \\
NGC3842-cE        &  $>$ 15.00              & $-$0.17$^{+0.04}_{-0.05}$ &    0.09$^{+0.12}_{-0.10}$ && 12.3$\pm$\,3.6&    0.19$\pm$\,0.04 &  2.0 && 12.7$\pm$\,3.7&    0.18$\pm$\,0.04   \\
Perseus-cE1       &  $>$ 15.00              & $-$0.20$^{+0.10}_{-0.15}$ &    0.22$^{+0.10}_{-0.10}$ &&  8.9$\pm$\,2.3&    0.02$\pm$\,0.03 &  4.1 && 13.0$\pm$\,3.4&    0.17$\pm$\,0.03   \\
\midrule                                                                                                                                                  
NGC2892-cE        & 12.51$^{+1.99}_{-2.51}$ & $-$0.06$^{+0.15}_{-0.08}$ &    0.14$^{+0.02}_{-0.07}$ && 12.8$\pm$\,4.1&    0.17$\pm$\,0.06 &  1.1 && 12.5$\pm$\,4.0&    0.03$\pm$\,0.06   \\
NGC1272-cE1       &  $>$ 15.00              & $-$0.14$^{+0.10}_{-0.15}$ &    0.38$^{+0.14}_{-0.29}$ && 10.6$\pm$\,2.5&    0.20$\pm$\,0.02 &  3.9 && 11.5$\pm$\,2.7&    0.02$\pm$\,0.02   \\
PGC038205         &  $>$ 15.00              & $-$0.30$^{+0.10}_{-0.05}$ &    0.33$^{+0.03}_{-0.03}$ && 12.9$\pm$\,2.3&    0.18$\pm$\,0.02 &  1.4 && 11.9$\pm$\,2.1&    0.16$\pm$\,0.02   \\
\midrule                                                                                                                                                               
CZcE44            &  $>$ 15.00              & $-$0.50$^{+0.17}_{-0.16}$ &    0.14$^{+0.14}_{-0.20}$ && 10.5$\pm$\,3.7&    0.04$\pm$\,0.07 &  1.7 && 11.0$\pm$\,3.8&    0.02$\pm$\,0.07   \\
CZcE95            &  $>$ 15.00              & $-$0.23$^{+0.16}_{-0.14}$ &    0.55$^{+0.22}_{-0.21}$ && 11.4$\pm$\,3.9&    0.03$\pm$\,0.07 &  2.8 && 11.5$\pm$\,3.9&    0.02$\pm$\,0.07   \\
PGC050564-cE2     &  3.36$^{+3.64}_{-1.36}$ &    0.32$^{+0.07}_{-0.12}$ &    0.21$^{+0.18}_{-0.17}$ && 10.8$\pm$\,3.4& $-$0.00$\pm$\,0.05 &  3.7 && 10.6$\pm$\,3.3& $-$0.02$\pm$\,0.05   \\
AHcE1             &  4.75$^{+0.55}_{-1.59}$ &    0.29$^{+0.04}_{-0.06}$ &    0.14$^{+0.02}_{-0.02}$ &&  9.7$\pm$\,2.8&    0.20$\pm$\,0.04 &  3.5 &&  9.6$\pm$\,2.8&    0.20$\pm$\,0.04   \\
AHcE2             &  3.41$^{+1.69}_{-0.21}$ &    0.15$^{+0.06}_{-0.04}$ &    0.12$^{+0.13}_{-0.13}$ && 10.8$\pm$\,3.6&    0.19$\pm$\,0.06 &  0.5 &&  9.0$\pm$\,3.0&    0.17$\pm$\,0.06   \\
CZcE57            &  $>$ 15.00              & $-$0.22$^{+0.09}_{-0.09}$ &    0.26$^{+0.01}_{-0.01}$ && 12.8$\pm$\,3.0&    0.20$\pm$\,0.01 &  1.0 && 11.8$\pm$\,2.8&    0.16$\pm$\,0.01   \\
CZcE194           & 10.40$^{+5.40}_{-4.80}$ &    0.01$^{+0.08}_{-0.10}$ &    0.12$^{+0.18}_{-0.12}$ && 11.9$\pm$\,3.5&    0.23$\pm$\,0.04 &  3.6 && 11.7$\pm$\,3.5&    0.21$\pm$\,0.04   \\
NGC2970           &  1.29$^{+0.02}_{-0.02}$ & $-$0.07$^{+0.02}_{-0.04}$ & $-$0.11$^{+0.03}_{-0.03}$ &&  7.1$\pm$\,1.7&    0.11$\pm$\,0.01 &  2.8 &&  2.6$\pm$\,0.6&    0.09$\pm$\,0.01   \\
NGC5567-cE1       &  5.66$^{+0.69}_{-0.66}$ & $-$0.39$^{+0.10}_{-0.10}$ & $-$0.35$^{+0.03}_{-0.05}$ &&  8.3$\pm$\,2.8&    0.26$\pm$\,0.06 &  6.1 &&  6.0$\pm$\,2.0&    0.26$\pm$\,0.06   \\
NGC5567-cE2       &  $>$ 15.00              & $-$0.22$^{+0.05}_{-0.09}$ &    0.12$^{+0.03}_{-0.02}$ && 11.7$\pm$\,2.9&    0.16$\pm$\,0.02 &  1.9 && 11.0$\pm$\,2.7&    0.16$\pm$\,0.02   \\
VCC165-cE         &  3.11$^{+1.09}_{-0.61}$ & $-$0.16$^{+0.05}_{-0.05}$ &    0.01$^{+0.15}_{-0.17}$ &&  7.8$\pm$\,2.6&    0.10$\pm$\,0.07 &  1.4 &&  3.5$\pm$\,1.2&    0.07$\pm$\,0.07   \\
\bottomrule
\end{tabular}

\vspace{0.2cm}
{Stellar population properties from both line indices and full-spectral-fitting approaches. Mean SSP ages, metallicities and [$\alpha$/Fe] ratios from the line index approach are described in Appendix B. The mean mass- and luminosity-weighted ages and total metallicities are obtained from the SFHs using {\tt STARLIGHT}.  T50 (the time the galaxy needed to build up half of its mass) is derived from the mass-weighted SFH and used to derive the [$\alpha$/Fe] ratios in those galaxies where the line index measurements returned unreliable results (marked with a \ddag). The different parts of the table correspond to the four visual classes as in previous tables (\textit{no host, near host, within halo} and \textit{disrupted})}
\end{table*}

\begin{table}\centering
\label{table:4}                      
\caption{Stellar masses}    
\begin{tabular}{c | c c c c}   
\toprule
\textbf{Galaxy}     & \textbf{ $\mathrm{M_{*,index}}$ }  & \textbf{ $\mathrm{M_{*,L}}$ } & \textbf{ $\mathrm{M_{*,M}}$ } & \textbf{ $\mathrm{M_{*,SFH}}$ }\\  
                  &  \multicolumn{4}{c}{($\times$10$^{9}$M$_{\odot}$) } \\
\midrule		
\midrule		 
AHcE0              &   6.4  &   8.7 &   6.9 &    7.3	\\
CGCG-036-042       &   4.2  &   3.1 &   4.1 &    2.6	\\
J16+14             &   2.3  &   2.9 &   3.2 &    2.6	\\
J07+50             &   3.1  &   3.9 &   4.4 &    3.2	\\
PGC012519          &  11.2  &  14.6 &  15.4 &   13.9  \\
\midrule     
LEDA\,3126625      &   2.5  &   2.2 &   2.4 &    2.1	\\
LEDA\,4544863      &   1.4  &   1.4 &   1.5 &    1.2	\\
NGC5846-cE         &   1.5  &   1.7 &   1.8 &    1.7	\\
NGC2832-cE         &   2.7  &   4.0 &   4.2 &    3.7	\\
NGC3842-cE         &   2.8  &   3.4 &   3.4 &    3.1	\\
Perseus-cE1        &   1.4  &   2.5 &   1.9 &    1.5	\\
\midrule   
NGC2892-cE ($\diamond$)      &   3.9  &   3.9 &   3.9 &    3.9	\\
NGC1272-cE1 ($\diamond$)     &   1.9  &   1.9 &   1.9 &    1.9	\\
PGC038205 ($\diamond$)       &   5.7  &   5.7 &   5.7 &    5.7	\\
\midrule   
CZcE44             &   9.2  &  15.1 &  14.3 &   12.9	\\
CZcE95(\dag)       &   8.6  &   9.9 &   9.9 &    8.8	\\
PGC050564-cE2(\dag)&   4.1  &   8.5 &   8.9 &    7.7	\\
AHcE1              &   4.4  &   7.1 &   7.1 &    6.9	\\
AHcE2              &   3.2  &   3.3 &   3.8 &    3.1	\\
CZcE57(\dag)       &  20.0  &  23.8 &  26.8 &   22.9	\\
CZcE194            &   4.5  &   5.4 &   5.6 &    4.9	\\
NGC2970            &   1.3  &   2.5 &   5.4 &    2.1	\\
NGC5567-cE1        &   3.1  &   4.9 &   6.4 &    4.5	\\
NGC5567-cE2(\dag)  &  17.9  &  20.0 &  20.7 &   17.4	\\
VCC165-cE          &   3.9  &   5.5 &  10.7 &    4.3	\\
\bottomrule	                            
\end{tabular}

\vspace{0.2cm}
{Stellar masses of the sample. They have been derived from the different M/L obtained from both the line indices approach ($\mathrm{M_{*,index}}$) and the full-spectral-fitting. In the latter, three estimates are shown, one considering all the SSP that contribute to the SFH ($\mathrm{M_{*,SFH}}$) and those from the mean mass/light- weighted SSP value ($\mathrm{M_{*,L}}$ and $\mathrm{M_{*,M}}$). For the galaxies marked with a $\diamond$, we use the previously published stellar mass from J+16 to overcome the high uncertainties in the magnitudes retrieved from SDSS (see Appendix A). In addition, those marked with a \dag  correspond to the four cEs with high probabilities of having an overestimated size and thus all their masses could decrease by up to 0.7\,dex}. The different parts of the table correspond to the four visual classes as in previous tables (\textit{no host, near host, within halo} and \textit{disrupted}).
\end{table}

\subsection{Stellar populations, SFHs and stellar masses}
We study the stellar populations on the basis of both line indices and full spectral fitting approaches. The advantage of the full spectral fitting is that it provides mass-weighted estimates of the age and the metallicity. If there are any recent events of star formation, which would outshine the older populations in the index approach, the stellar populations obtained from the full-spectral-fitting better represent the bulk of the stars. In addition, this technique provides the history of how the galaxy was (or is being) formed, by tracing its fossil imprint at different epochs. The full-spectral-fitting approach has proven to render reliable results (e.g. \citealt{CidFernandes2005}; \citealt{Koleva2008}; \citealt{Ferre-Mateu2014}; \citealt{GonzalezDelgado2014}) and it further breaks the strong age-metallicity degeneracy that line indices are so much affected by \citep{Sanchez-Blazquez2011}. 

The Single-Stellar Population models (SSP) used in both approaches are the newest extension of the {\tt MILES} SSPs (\citealt{Vazdekis2015}; \citealt{Vazdekis2016}), which cover a wide range of ages (0.03 to 14.5\,Gyr), total metallicities ($-$2.27 to +0.40\,dex) and different IMF slopes and shapes. We are using the scaled solar models and then applied an empirical proxy to obtain the [$\alpha$/Fe] abundances  (see Appendix B). While using a non-universal IMF has been proven to have an important impact when deriving stellar populations of massive galaxies (i.e. a bottom-heavy IMF would increase the stellar masses by a factor of $\sim$2), its impact on lower mass galaxies is very mild (e.g. \citealt{Ferre-Mateu2013}). Therefore we assume a universal Kroupa IMF for this exercise. We obtain an estimate for the ages and metallicities using H$\beta_{o}$ \citep{Cervantes2009} as our main age-sensitive index and [Mg/Fe]$^{\prime}$ \citep{Thomas2003} as the total metallicity one, quoted in Table 3. Appendix B presents a full description of the line index technique and how ages, metallicities and $\alpha$-enhancements are derived, together with comparisons from the literature. 

We then use the full spectral fitting code {\tt STARLIGHT} \citep{CidFernandes2005}, creating a combination of SSP model predictions that best resembles each spectrum. Although the SSP models cover from the UV to the IR region, we fit the same spectral region as in the stellar kinematics (3800 to 7400\,\AA\,). Using the entire SDSS spectral range renders ages typically $\sim$5\% older but with worse fitting parameters (e.g. higher $\chi^{2}$). Figure 8 is an example illustrating the methodology used. It shows the fits for both the entire SDSS spectral range and for the narrower one in a cE spectrum of our sample. The inset shows the spectral range relevant for the line index approach (used in Appendix B).

One has to be careful when interpreting the derived SFHs. Rather than taking each individual burst of formation, one should consider averaged episodes of star formation, i.e. divide the history into young, intermediate and old episodes, for example. This will account for the uncertainties associated to the full-spectral-fitting procedure while providing a robust sense of the galaxy SFH. We compute the associated errors in the mean ages and metallicities as in \citet{LaBarbera2010}, which are on average $\sim$20\% on the ages and $\sim$0.05\,dex for the metallicities. Another way to interpret such SFHs is by inferring the cumulative stellar mass, which represents how fast/slow the galaxy built up its stellar mass. 

Figure 9 shows the derived SFH for each individual cE and its cumulative stellar mass fraction. Each column represents a visual class and environment has been colour-coded. This sample of galaxies shows a variety of SFHs, with a slight dependence on environment. Field galaxies show predominantly slow and more extended formation timescales, and typically started later in time. Group and cluster galaxies show a wider variety of timescales, with those in the centres of the clusters or close to the cluster brightest galaxy showing an earlier and faster build up of their stellar mass. Those in the cluster outskirts tend to have SFHs that instead resemble more the field type, with very extended SFHs that peaked 8\,-\,10\,Gyr ago. We leave the discussion about the relation with the visual classification for Section 4.1, when all discriminant tools are analysed separately, and here we merely state the derived mean mass- and luminosity-weighted estimates from this approach in Table 3. 

Another interesting parameter is T50, which quantifies how long it took to build up half of the stellar mass of the galaxy. This parameter is directly calculated from the SFHs, quoted in Table 3 as the time elapsed since the Big Bang. For example, T50\,=\,2 means that the galaxy took 2\,Gyr to build up half of its stellar mass, even if it is unknown when it started. This parameter is important because it is connected to the enhancement of $\alpha$-elements in a galaxy, in the sense that galaxies which exhibit enhanced ratios typically form on very short timescales \citep{Thomas2005} and thus have shorter T50s. Larger T50 values are more indicative of slower and extended star formation histories. Therefore, if we know the $\alpha$ abundances of our galaxies, we can obtain another estimate for the galaxy formation timescales. Here we use the empirical relation of \citet{delaRosa2011}, as it covers a range in [$\alpha$/Fe] that is more representative to the cEs in our sample (see also J+16). 

The stellar mass is obtained by adopting the M/L from the stellar population fits and converting the galaxy magnitude into luminosity. As we used different approaches for the stellar population analysis, we now have four estimates for the M/L, and hence four estimates for the stellar mass, as quoted in Table 4. First, there is the $\mathrm{M_{*,index}}$, the one derived from the SSP age and metallicity measured with the line indices. But because we now have the true star formation history of each galaxy, we also have the M/L that comes from each individual SSP contributing into reproducing the galaxy SFH ($\mathrm{M_{*,SFH}}$). Furthermore, we also obtain a mean mass-weighted age and metallicity value ($\mathrm{M_{*,M}}$), and a luminosity-weighted estimate ($\mathrm{M_{*,L}}$). Although ($\mathrm{M_{*,SFH}}$) is the more realistic estimate, both the $\mathrm{M_{*,L}}$ and the $\mathrm{M_{*,index}}$ are a better approximation to the SED fitting that is usually employed in the literature and thus will be the ones used hereafter. As previously discussed, some of the galaxies have overestimated SDSS sizes, which translates into higher stellar and dynamical masses. While such variations are not significant for most of the objects, figure 15 shows that NGC\,1272-cE1, NGC\,2892-cE and PGC\,038205 should be considered differently. Therefore we will use the published stellar masses from J+16 for these three objects. In addition, the four galaxies for which smaller sizes (and stellar masses) are expected (Figure 16) are also flagged in Table 4. 

Figure 10 compares the dynamical mass calculated in section 3.1 with each one of the derived stellar masses obtained here. While the different estimates vary by a factor of $\sim$0.6\,dex, the general trend remains unchanged. Most of the points are close to the 1:1 relation, but there are a few galaxies showing larger deviations in all the relations. This implies high dynamical-to-mass ratios, as will be discussed in the next section.

\section{Discussion}
It has been proposed that cEs can be differentiated into two type of origins: an intrinsic one (nature; e.g. \citealt{Wirth1984}; \citealt{Kormendy2009}; \citealt{Kormendy2012}; \citealt{Huxor2013}; \citealt{Paudel2014b}); or a stripped origin (nurture; e.g. \citealt{Faber1973}; \citealt{Bekki2001}; \citealt{Huxor2011}; \citealt{Pfeffer2013}). Different predictions for properties of such cEs are expected for each case. The stripping process will leave the properties of the core relatively untouched while removing a large fraction of the galaxy's stellar mass. This means that the compact remnants should have stellar populations, velocity dispersions and black hole masses similar to those in the core of the progenitor galaxy, but with lower stellar masses and smaller sizes and thus deviating from some of the local scaling relations (e.g. \citealt{Graham2002}; \citealt{Chilingarian2009}). If we are lucky enough to see a galaxy being currently stripped or with signs of such interaction, the stellar populations should also reflect such structural changes. However, if cEs are instead the low-mass end of ETGs, then we would expect them to follow most of the local scaling relations and have stellar populations intrinsic to their family. We remind the reader again that by low-mass ETGs we are referring to compact objects with high densities and stellar masses in the 10$^{8}$-10$^{10}$M$_{\odot}$ range, following the mass-size relation described by bright ellipticals alone (e.g. \citealt{Brodie2011}; \citealt{Misgeld2011}).

\begin{figure*}
\centering
\includegraphics[scale=0.6]{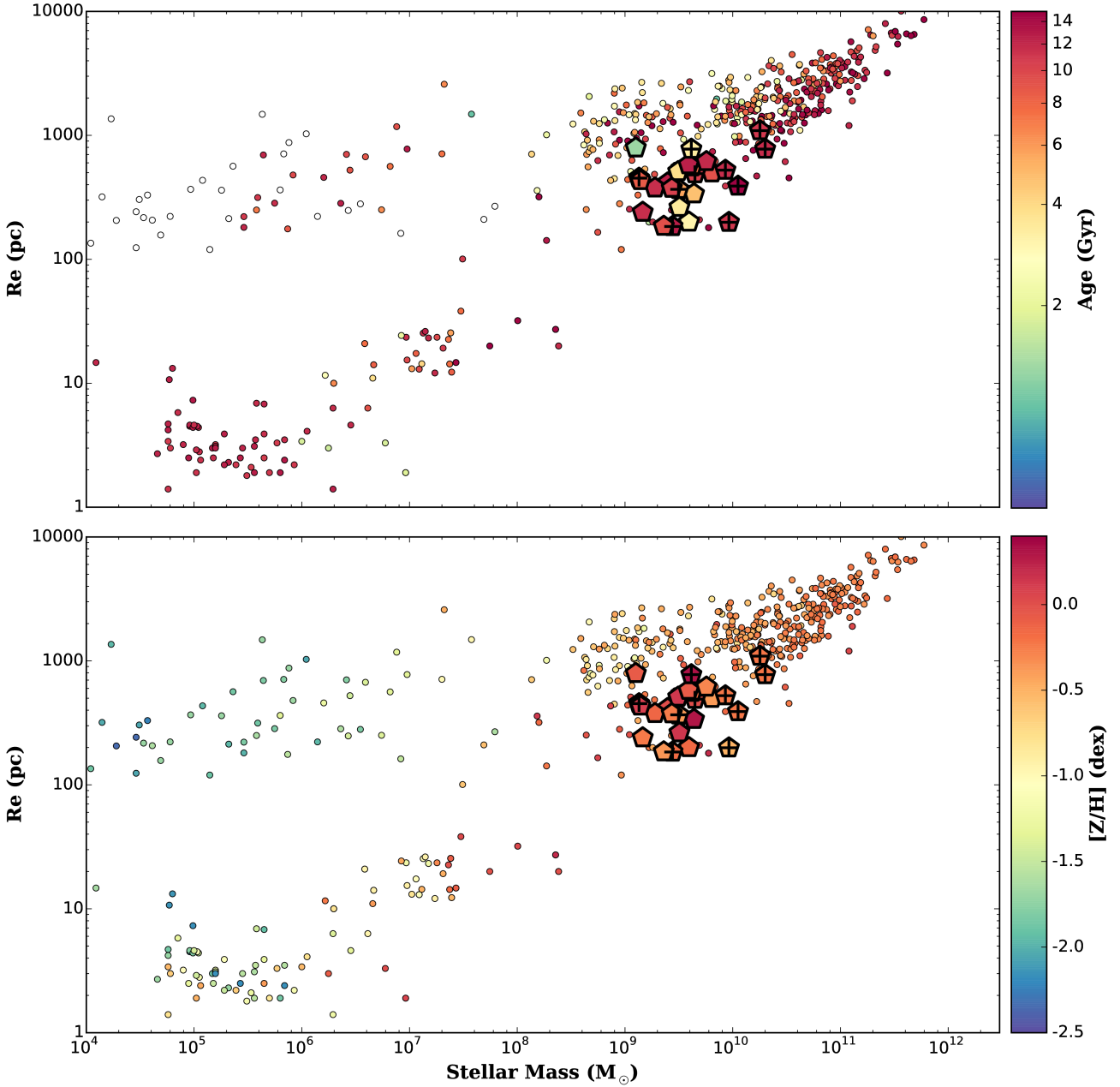}\\
\caption{Stellar mass \textit{vs} size relation. The stellar mass-size relation for all systems in AIMSS (N+14; circles) is shown. Those with ages and metallicities from J+16 have been colour-coded accordingly to these parameters, together with our sample of cEs (colored pentagons). We use here the stellar masses, ages and metallicities derived from the line indices, to be consistent with the literature. Galaxies with black crosses represent those with sizes from the SDSS, and therefore should be considered as upper limits.}
\label{figure:11}
\end{figure*}

\begin{figure*}
\centering
\includegraphics[scale=0.65]{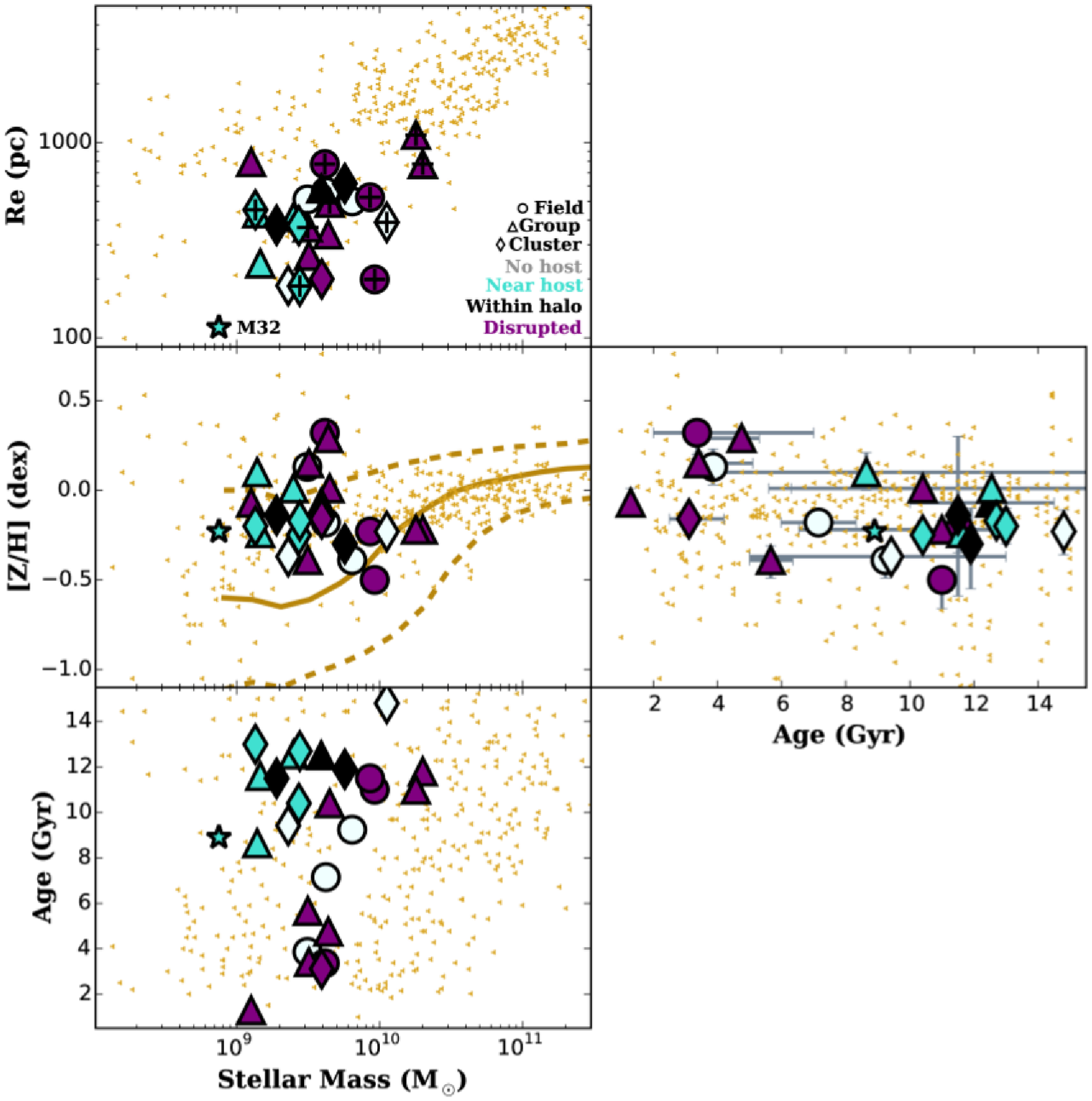}\\
\caption{Scaling relations of compact systems. The prototype of cE, M32, is shown in all panels as a cyan star. \textit{Top-left}: Mass-size relation from Figure 11 focused on the 10$^{8}$ - 10$^{11}$M$_{\odot}$ mass region. Yellow symbols correspond to the objects from the AIMSS sample whereas our sample of cEs are represented by different symbols according to their environment and colour-coded by their visual classification. \textit{Middle-left}: Mass-metallicity relation from \citet{Gallazzi2005} (yellow line) and its intrinsic scatter (dashed yellow lines). The systems from J+16 (AIMSS objects with stellar population parameters) have been shifted 0.15\,dex up to account for aperture effects (SDSS fiber size \textit{vs} longslit aperture). The stellar populations used in this figure have been derived under the line indices approach for comparison with the literature. \textit{Bottom-left}: Mass-age relation for our objects and the J+16 sample. This relation shows a huge scatter at lower masses ($<$5$\times$10$^{9}$M$_{\odot}$), in particular for the \textit{disrupted} class. \textit{Middle-right:} Age-metallicity relation. It is seen that those galaxies with intermediate to young ages and solar or over-solar metallicities, are more prone to be of the \textit{disrupted} class, typically embedded in streams. The rest of the \textit{disrupted} galaxies are older with mildly sub-solar metallicities. The same region is populated by the galaxies \textit{near host} and \textit{within halo}. Galaxies that have \textit{no host} show intermediate to old ages and moderately low metallicities (in agreement with the mass-metallicity scaling relation).}
\label{figure:12}
\end{figure*}

\subsection{Results from the individual discriminant tools}
In the following, we discuss the different discriminant tools separately (structural parameters, stellar populations, SFHs and mass ratios) for each class of cE. We try to determine the most plausible origin, evolutionary stage and even progenitor type, if possible. Combining the information from all the discriminants, it will be then decided which is the most outcome for each individual cE, summarized in Table 5 at the end of the section. For each discriminant tool, we start by analyzing the properties of those objects for which the visual inspection alone already confirms their tidal stripping origin, i.e. the 11 \textit{disrupted} galaxies. Although their origin is known, we additionally try to determine both their evolutionary stages and progenitors. We then compare their derived properties with those in galaxies for which the visual inspection alone is not enough, i.e. the 3 galaxies \textit{within halo} and the 6 \textit{near host}. With this, we will be able to confirm if the stripping process took place and again, which stage and progenitor are more plausible. Finally, for the 5 galaxies with \textit{no host}, we will see if their properties are more compatible with being a stripped object that ran away from its host (thus, similar to the galaxies with a host that are the confirmed end products of a threshing) or if they rather show properties more similar to low-mass ETGs. 

We start with the structural parameters, with Figure 11 showing the loci of our cEs within the well-known mass-size relation, where galaxies get systematically smaller as we move towards lower stellar masses. This relation is known to follow a power-law only for the ETG regime and is then truncated to follow the dE and dSph (dwarf spheroidal) families. However, systems as cEs and UCDs seem to follow the relation of ellipticals, with much smaller sizes than their similar mass dE counterparts (e.g. \citealt{Brodie2011}; \citealt{Misgeld2011}). This is represented by the open circles, which correspond to objects from the AIMSS catalog (N+14) throughout the entire mass range. Both panels show those AIMSS objects for which a stellar population analysis was performed by J+16 and our cE sample (pentagons), colour coded according to their stellar populations parameters. It is seen that our objects have similar properties to other objects close to their loci in the mass-size plane, with typically intermediate to old ages and roughly solar total metallicities. All but two of the objects (the most massive ones) are populating the areas corresponding to high densities, close to the zone of avoidance described in N+14. This confirms their nature as cEs. Note that to be consistent with the literature, we use the values from the line index approach here. 

Next, we investigate how the different classes are represented in this scaling relation. Figure 12 (top-left panel) shows the same mass-size relation but centered on the cE regime, with symbols representing the different environments of the galaxies and colours for the four visual classifications. The cyan star represents the prototype of cEs, M32 (values from N+14). The most massive objects of our sample are two galaxies of the \textit{disrupted} class (NGC\,5567-cE2 and CZcE57), which also show the largest sizes in the sample. This could indicate that they are in an early stage of stripping and thus still have to experience structural changes, with further evolution towards smaller sizes and stellar masses. Despite having flagged unreliable sizes, such an early stage is still plausible even if considering a smaller size and thus a smaller stellar mass (see Figure 16). In that case, they would move closer to the rest of galaxies but still would be the most massive ones. The rest of the \textit{disrupted} galaxies cover the intermediate regions of stellar masses (3-9\,$\times$10$^{9}$M$_{\odot}$), which could indicate either intermediate (if they have larger sizes) or later stages (for the smaller ones) in the stripping process. The two other galaxies with unreliable sizes are in this aregime and a smaller size would move them towards a later stage. The same region is populated by the galaxies with \textit{no host}. As these galaxies seem to follow the elliptical mass-size relation, the intrinsic origin is plausible for all them, but this tool alone is not sufficient to even speculate about it. Those \textit{near host} but with no signs of interaction tend to occupy the low-mass and small size end of the relation, compatible with being the remnant product once the tidal stripping is completed. The galaxies classified as \textit{within halo} populate intermediate masses and sizes, thus they could be both the end product (NGC\,1272-cE1) or the beginning of it (NGC\,2892-cE and PGC\,038205). Overall, further insight from another discriminant is required. 

To this end, middle- and bottom-left panels of Figure 12 show the SDSS mass-metallicity and the mass-age relations, respectively. The mass-metallicity one is determined by the SDSS relation from \citet{Gallazzi2005} (yellow line). Note the J+16 points (yellow symbols) have been arbitrarily shifted 0.15\,dex up to account for aperture corrections and we have substituted the ages that were at the limit of the model grids in the line indices for their mean luminosity weighted value. This is a robust assumption as the ages in those cases are all compatible with being old in all the approaches ($>$10\,Gyr, see Figure 17). We remind the reader about the caveat related to the galaxy sizes and aperture effects within our own sample. Those galaxies covering more than 1.5\,$\mathrm{R_e}$ will have slightly lower metallicities and be older than those of $\sim$1$\mathrm{R_e}$ and conversely,, galaxies with less than 0.5\,$\mathrm{R_e}$ will be more metal-rich and younger than the rest, as we discuss below. 

In both panels a difference in the stellar properties of the objects is seen. Galaxies more massive than $\sim$5\,$\times$10$^{9}$M$_{\odot}$ follow the mass-metallicity relation tightly and show uniformly old stellar ages. For the \textit{disrupted} galaxies in that region, this represents an early stage where the stripping has just commenced. A lower stellar mass for NGC\,5567-cE2, CZcE57 and CZcE95 would again reinforce their early stage even more, as they would fall directly on the relation. For the galaxies with \textit{no host} in that regime (PGC\,012519 and AHcE0), an early makes no sense, and instead they are more compatible with simply being the low-mass ETGs. This is also in agreement with the intermediate to old ages they show. Interestingly, the \textit{within halo} galaxy PGC\,038205 also occupies this area. In this case, an intrinsic origin as a low-mass galaxy could also be possible, as seen from cosmological simulations \citep{Martinovic2017}. However, more information from the other diagnostic tools is needed to confirm this. 

Galaxies below $\sim$5\,$\times$10$^{9}$M$_{\odot}$ show a larger scatter in both panels. The objects in this mass regime tend to have higher metallicities than expected but with a spread in ages. Under the assumption of a stripping process, we expect that the higher the deviation from the mass-metallicity relation, the more advanced is the stripping stage. For those \textit{near host} this directly implies they are the completed stage of stripping, which is in agreement with the location of M32 (cyan star, with stellar populations from \citet{Worthey2004}). However, a very late stage of stripping could also be depicted for the \textit{disrupted} galaxies AHcE1, AHcE2 and PGC\,050564-cE2. Interestingly, all three galaxies have very large physical coverages. This means that they would be even more metal-rich, further reinforcing a late stage of stripping. Similarly, the \textit{no host} galaxy J16+14 presents abnormally high metallicities, pointing out that this would be, instead, a run away stripped galaxy. The rest of \textit{disrupted} galaxies in this mass regime, due to their slightly lower deviations, would be more compatible with an intermediate to late stage of stripping (CZcE194, NGC\,2970 and VCC165-cE), similar to the ones \textit{within halo}. The mass-age relation shows that \textit{disrupted} galaxies in this mass regime show typically young ages ($<$6\,Gyr), whereas all the \textit{near host} galaxies, NGC3842-cE1 and NGC\,1272-cE1 (\textit{within halo}) show basically old stellar populations. 

However, high metallicities can also be achieved if the galaxy suffered other type of interactions; e.g. mergers, strangulation after it quenched or encounters with gas-rich galaxies (e.g. \citealt{Emsellem2008}; \citealt{Chilingarian2009}; \citealt{Peng2015}). The only way to differentiate the stripping mechanism from these ones is with the stellar ages of the cEs. New star formation will bias the luminosity-weighted stellar age towards younger ages. Those with younger ages will thus represent progenitors that had a reservoir of cold gas, creating new stars while the stripping is underway. Those purely old would better be represented by ETGs progenitors whose outer stars were simply threshed, with no gas involved. The right panel of Figure 12 shows the age-metallicity relation, where the dichotomy seen for the \textit{disrupted} galaxies is emphasized. About half of them have solar or above solar metallicities and show younger ages ($<t>$= 3.5\,Gyr and $<$[Z/H]$>$ = 0.0\,dex). The fact that we are seeing the stripping actually happening from the visual inspection, an scenario where a spiral-like progenitor is at latter stages of its stripping, fits better. The exception is NGC\,5567-cE1, which shows a sub-solar metallicity that is compatible with its stellar mass. While the progenitor would still be compatible with being a spiral-like one, such metallicity is more similar to what is seen for the rest of the \textit{disrupted} galaxies -- that is, more representative of an early stage of tidal stripping. This second group of \textit{disrupted} galaxies shows on average old ages ($<t>$= 11.0\,Gyr), solar or mildly sub-solar metallicities ($<$[Z/H]$>$ = $-$0.2\,dex) and their progenitors would be somewhat intermediate mass ETGs that just started being stripped. CZcE44 has also a low metallicity similar to the \textit{no host} galaxies and it also follows the mass-metallicity relation. This would be the case of an intermediate-mass, already rather compact ETG that has just started being stripped, compatible with its visual image that shows some signs of disruption.

This region is also occupied by all \textit{within halo} galaxies. If they are indeed the end product of the stripping process, as the previous properties indicate, such old ages and moderate metallicities imply ETG-like progenitors of intermediate mass ($\sim$2-3\,$\times$10$^{10}$\,M$_{\odot}$). The location of M32 is very similar to this class of cEs through all the panels, further reinforcing their stripped origin. For the three galaxies \textit{within halo}, also in the same region, this discriminant is not strong enough, as they could represent either early stages (such as the \textit{disrupted} type) or completed ones (such as \textit{near host} ones). Their old ages and moderately sub-solar metallicities could also indicate an intermediate-mass ETG progenitor. We finally study the galaxies with \textit{no host}. Three are located in the center of the relation, between the two areas defined by the disrupting class. These \textit{no host} cEs show intermediate ages and sub-solar metallicities ($< t >$= 9.0\,Gyr and $<[Z/H]>$ = $-$0.3\,dex). While CGCG-036-042 is clearly differentiated and compatible with being a low mass, unstripped ETGs, AHcE0 and J07+50 are more similar to the end products of stripping. In addition, J16+14 is located at the region of young ages and high metallicities, again pointing out that it might be one of the run-away systems proposed by \citet{Chilingarian2015}. The other interesting galaxy, at the other end of the age-metallicity relation, is PGC\,012519. This galaxy follows the mass-metallicity relation, although is a bit of an outlier in the mass-size one and also occupies a mass-age position different to any other galaxy. But, as it happens with the other cEs, we need the information from the other discriminants to better understand its origin.  

Appendix B shows similar scaling relations of mass and age with the derived $\alpha$-enhancements. While there is a small hint of galaxies being more enhanced at older ages, there is no trend with stellar mass, therefore this is not useful as a discriminant. The reader is referred to Figure 19 of the Appendix if more information about this behavior is required. 

\begin{figure*}
\centering
\includegraphics[scale=0.5]{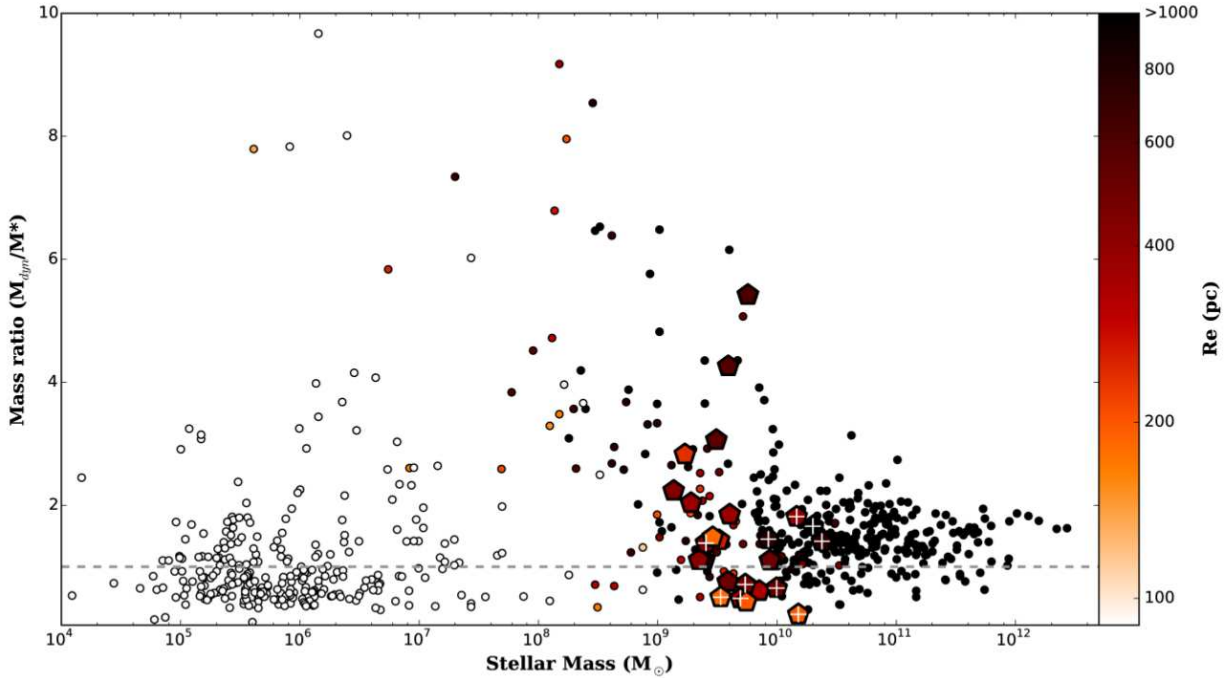}\\
\caption{Dynamical-to-stellar mass ratio \textit{vs} stellar mass. Small dots are all the systems of N+14 whereas large pentagons are the sample of cEs studied here. All points have been also colour-coded by size. Note that it has been limited to the size range of our cEs (100$\le\mathrm{R_e}\le$\,1000\,pc). Therefore, galaxies with less than 100\,pc appear as white circles and galaxies larger than 1000\,pc appear as black ones. White crosses mark those galaxies with size estimates from the SDSS instead of published values.}
\label{figure:13}
\end{figure*}

\begin{figure}
\centering
\includegraphics[scale=0.6]{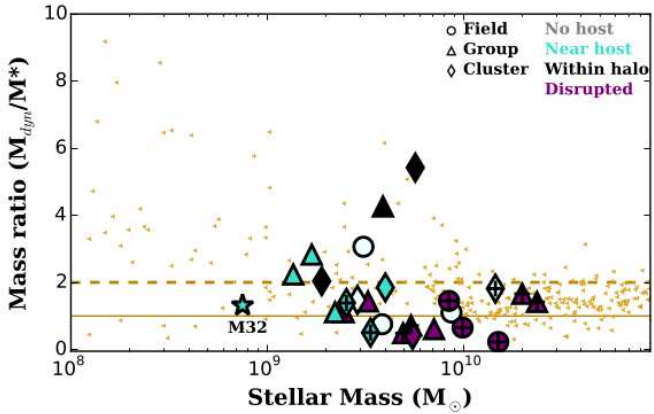}\\
\caption{Mass ratio \textit{vs} stellar mass, colour-coded by the different visual classes and with symbols representing the environment. The solid yellow line marks the unity, while the dashed one marks the M$_{\mathrm{dyn}}$/M$_{*}$ > 2 limit. M32 is shown again for comparison, being almost at unity. This discriminant has turned out to be weakest one, but it can help secure the stage for those with large deviations from unity.}
\label{figure:14}
\end{figure}

Therefore, we go back to the derived SFHs in Figure 9. In most of the cases, these should be able to provide crucial information about the progenitor type, but also shed some more light onto the nature of the \textit{no host} galaxies. For the \textit{disrupted} galaxies, we see two types of SFH. Some galaxies show secondary bursts of formation at recent epochs, which are the reason for their younger ages in the SSP estimates. Such posterior events of star formation can only be explained if there is a cold gas reservoir in the progenitor galaxy that boosts new star formation when the stripping event occurs. Therefore, such cEs are thought to have a spiral-like progenitor. Not much can be said about the exact stage in the stripping process, only that it occurred recently (last 2\,Gyr). This is the case for VCC165-cE, AHcE2, NGC\,2970 and PGC\,050564-cE, which all have high metallicities. 

High metallicities can also be explained if the progenitor galaxy was quenched by strangulation and it has was posteriorly threshed. In this case, the SFH would show an early and peaked event, representative of an ETG-like progenitor, followed by an extended episode of formation at very low rates that can persist for at least 4\,Gyr after quenching. This strangulation mechanism has been shown to be the most efficient mechanism for shutting down star formation in galaxies with masses below 10$^{11}$\,M$_{\odot}$ \citep{Peng2015}. Again, this tool will not allow us to determine the evolutionary stage, but we can see that the stripping event started recently for at least CZcE57 and NGC\,5567-cE2 (the low rate of star formation goes down to current time). For AHcE1, CZcE44, CZcE95 and CzCE194, the event could have at started anytime after the galaxy was quenched around 6\,Gyr ago. NGC\,5567-cE1 is the only galaxy in this class that shows a completely different SFH. This one resembles to the extended SFHs of the majority of the \textit{no host} galaxies. Such types of SFH have been found for low-mass ETGs (e.g. \citealt{Thomas2005}; \citealt{Ferre-Mateu2013}). Therefore, NGC\,5567-cE1 would be an intrinsically low-mass ETG that has started to be stripped by its host very recently, and the three \textit{no host} galaxies with such similar SFHs, would be, indeed, of the intrinsic origin. Only PGC\,012519 shows a completely different SFH, with an early and peaked formation that resembles its more massive counterparts (see next section). The formation of J07+50 also seems to happen earlier than the rest of galaxies with \textit{no host}. This can be attributed to their environment, that triggers star formation earlier, as both cEs are located in the outskirts of clusters. 

We look at the SFHs of those with a host but without signs of interaction. By now, it is clear that they are compatible with being the end product of a tidal stripping event. Comparing with the SFHs in the other classes, we find that LEDA\,3126625 and Perseus-cE1 would be compatible with a spiral-like progenitor, while the rest are more compatible with an ETG-like one. The stripping episode for LEDA\,4544863 and NGC\,2832-cE would have ended recently, but it cannot be determined for NGC\,5846-cE, NGC\,3842-cE, NGC\,1272-cE1. PGC\,38205 and NGC\,2892-cE show the peaked plus low rate SFHs down to the present time, which would be compatible with just having started their stripping event rather than being an intrinsically low-mass ETG like J07+50.

Lastly, we investigate the dynamical-to-stellar mass ratios of the sample, as they have been reported to be abnormally elevated for many compact systems (e.g. \citealt{Mieske2013}; \citealt{Forbes2014}; \citealt{Janz2015}; \citealt{Janz2016b}). Figure 13 shows such mass ratios \textit{vs} the stellar mass (this time from the luminosity-weighted M/L) for all the compact systems in the AIMSS sample (N+14; filled dots) and our galaxies (pentagons). They have been colour-coded by galaxy size, limiting it to the range in size from our sample (100$\le\mathrm{R_e}\le$\,1000\,pc). Therefore, galaxies with less than 100\,pc appear as white circles and galaxies larger than 1000\,pc appear as black ones. In addition, galaxies with a cross are those with SDSS sizes, which are most likely to represent an upper limit and should be taken cautiously. However, as discussed in Appendix A, such galaxies do not change dramatically in this plot even if considering smaller sizes. The figure shows that the deviations extend throughout the stellar mass range, although the most extreme cases are in the 10$^6$-10$^9$\,M$_{\odot}$ mass range. In fact, it is seen that the galaxies with larger deviations, i.e. M$_{\mathrm{dyn}}$/M$_{*}>$ 2, correspond to the low mass ones of our sample ($\sim$10$^9$\,M$_{\odot}$), which is consistent with the \citet{Forbes2014} findings. 

\citet{Pfeffer2013} presented a set of cosmological simulations that predicted the evolution of compact objects under a tidal stripping scenario. Although those predictions are based on UCDs, we assume the stripping process for cEs follow similar steps (the only thing that varies is the type of progenitor galaxy and the stellar mass). The evolutionary tracks described in such simulations are an extremely fast process (Figure 7 in \citealt{Forbes2014}), but present a two-step process. When the progenitor galaxy starts to be stripped, it loses the majority of its stellar mass but both the size and velocity dispersion remain relatively constant. This initially decreases the M$_{*}$ but keeps the inferred M$_{\mathrm{dyn}}$ relatively constant, giving elevated mass ratios. Afterwards, the remnant shrinks with a limited amount of mass loss, lowering its inferred M$_{\mathrm{dyn}}$ and thus moving close to unity while restoring its equilibrium. Therefore, the location of the cEs in such evolutionary tracks can give further information about their stage of tidal stripping. 

From Figure 14 it is clear that this diagnostic tool is not as reliable as the previous ones, with the majority of our objects close to unity. M32 lies also very close to unity, as expected for the end products of stripping. However, this tool might help to better understand the evolutionary stage of those cases that strongly deviate from unity, i.e. M$_{\mathrm{dyn}}$/M$_{*}>$ 2. There are six galaxies in our sample with such characteristics. The most extreme cases are the two galaxies \textit{within halo} whose properties suggested they were on the beginning phases of stripping (both the stellar populations and SFHs). Such a scenario is further reinforced with this third tool, where they are compatible with being on the first step of the stripping process. The other \textit{within halo} galaxy, NGC\,1272-cE1 also lies outside the scatter but its location would rather point to an finished stage of stripping. The two other systems with M$_{\mathrm{dyn}}$/M$_{*}>$ 2, which are classified as \textit{near host}, would also be compatible with being an end product of tidal stripping (NGC\,5846-cE and LEDA\,4544863). Interestingly, the other galaxy with a large mass ratio is J16+14. This is the same \textit{no host} galaxy that has shown properties more compatible with being a run-away case of a stripped galaxy. 

\begin{table*}\centering
\label{table:5}                      
\caption{Origin, evolutionary stage and progenitor}    
\begin{tabular}{c | c c c c c c}   
\toprule
  \textbf{Galaxy}     & \textbf{Mass-size}   & \textbf{Mass-metallicity} & \textbf{Age-metallicity}   & \textbf{SFHs} &    \textbf{Mass ratio}     &  \textbf{Verdict}    \\  
                             &  \multicolumn{6}{c}{(origin/evol. stage/progenitor) } \\

\midrule
\midrule
AHcE0          &  -                & LMG           &  -              &  LMG       &  -   & \textbf{\textcolor{JungleGreen}{LMG}           } \\
CGCG-036-042   &  -                & -             &  LMG            &  LMG       &  -   & \textbf{\textcolor{JungleGreen}{LMG}           } \\
J16+14         &  -                & SG            &  SG/S           &  LMG       &  SG  & \textbf{\textcolor{BurntOrange}{SG/complete/S} } \\
J07+50         &  -                & -             &  -              &  LMG       &  -   & \textbf{\textcolor{JungleGreen}{LMG}           } \\
PGC012519      &  -                & LMG           &  LMG            &  -         &  -   & \textbf{\textcolor{JungleGreen}{LMG (RELIC)}           } \\
\midrule                                                                          
LEDA\,3126625  &  SG?/complete?/-  & SG/complete/- &  SG/complete/E? & SG/-/S     &  -   & \textbf{\textcolor{JungleGreen}{SG/complete}\textcolor{BurntOrange}{S}  }\\
LEDA\,4544863  &  SG?/complete?/-  & SG/complete/- &  SG/complete/E? & SG/-/E     &  SG  & \textbf{\textcolor{JungleGreen}{SG/complete/E}                          } \\
NGC5846-cE     &  SG?/complete?/-  & SG/complete/- &  SG/complete/E? & SG/-/E     &  SG  & \textbf{\textcolor{JungleGreen}{SG/complete/E}                          } \\
NGC2832-cE     &  SG?/complete?/-  & SG/complete/- &  SG/complete/E? & SG/-/E     &  -   & \textbf{\textcolor{JungleGreen}{SG/complete/E}                          } \\
NGC3842-cE     &  SG?/complete?/-  & SG/complete/- &  SG/complete/E? & SG/-/E     &  -   & \textbf{\textcolor{JungleGreen}{SG/complete/E}                          } \\
Perseus-cE1    &  SG?/complete?/-  & SG/complete/- &  SG/complete/E? & SG/-/S     &  -   & \textbf{\textcolor{JungleGreen}{SG/complete/}\textcolor{BurntOrange}{S} } \\
\midrule                                          
NGC2892-cE     &  SG?/early?/-     & SG/early/-    &  SG/early?/E?   & SG/early/E & SG/early?/-  & \textbf{\textcolor{JungleGreen}{SG/early/E}    } \\
NGC1272-cE1    &  SG?/complete?/-  & SG/complete/- &  SG/early?/E?   & SG/-/E     & SG/late?/-   & \textbf{\textcolor{JungleGreen}{SG/}\textcolor{BurntOrange}{complete}\textcolor{JungleGreen}{/E} } \\
PGC038205      &  SG?/early?/-     & SG/complete/- &  SG/early?/E?   & SG/early/E & SG/early?/-  & \textbf{\textcolor{JungleGreen}{SG/}\textcolor{BurntOrange}{early}\textcolor{JungleGreen}{/E} } \\
\midrule                                                                             
CZcE44         &  SG/late?/-       & SG/early/-    &  SG/early/E     & SG/-/E     &  -   & \textbf{\textcolor{JungleGreen}{SG/}\textcolor{BurntOrange}{early}\textcolor{JungleGreen}{/E} }\\
CZcE95         &  SG/int?/-        & SG/early/-    &  SG/early/E     & SG/-/E     &  -   & \textbf{\textcolor{JungleGreen}{SG/early/E}                                                   }\\
PGC050564-cE2  &  SG/int?/-        & SG/late/-     &  SG/late/S      & SG/-/S     &  -   & \textbf{\textcolor{JungleGreen}{SG/late/S}                                                    }\\
AHcE1          &  SG/int?/-        & SG/late/-     &  SG/late/S      & SG/-/E     &  -   & \textbf{\textcolor{JungleGreen}{SG/late/}\textcolor{BurntOrange}{?}                           }\\                                          
AHcE2          &  SG/late?/-       & SG/late/-     &  SG/late/S      & SG/-/S     &  -   & \textbf{\textcolor{JungleGreen}{SG/late/S}                                                    }\\
CZcE57         &  SG/early?/-      & SG/early/-    &  SG/early/E     & SG/early/E &  -   & \textbf{\textcolor{JungleGreen}{SG/early/E}                                                   }\\
CZcE194        &  SG/int?/-        & SG/int-late/- &  SG/early/E     & SG/-/E     &  -   & \textbf{\textcolor{JungleGreen}{SG/}\textcolor{BurntOrange}{?}\textcolor{JungleGreen}{/E}     }\\
NGC2970        &  SG/int?/-        & SG/int-late/- &  SG/late/S      & SG/-/S     &  -   & \textbf{\textcolor{JungleGreen}{SG/late/S}                                                    }\\
NGC5567-cE1    &  SG/late?/-       & SG/early?/-   &  SG/early/S     & SG/early/LMG &  -   & \textbf{\textcolor{JungleGreen}{SG/}\textcolor{BurntOrange}{early/LMG}                          }\\
NGC5567-cE2    &  SG/early?/-      & SG/early/-    &  SG/early/E     & SG/early/E &  -   & \textbf{\textcolor{JungleGreen}{SG/early/E}                                                   }\\
VCC165-cE      &  SG/late?/-       & SG/int-late/- &  SG/late/S      & SG/-/S     &  -   & \textbf{\textcolor{JungleGreen}{SG/late/S}                                                    }\\
\bottomrule	                            
\end{tabular}

\vspace{0.2cm}
{Summary table for the discriminant tools used in this work to determine the origin and evolutionary stage for our sample of cEs. Again, the different parts of the table correspond to the four visual classes as in previous tables (\textit{no host, near host, within halo} and \textit{disrupted}). LMG stands for Low-Mass-ETG (thus having an intrinsic origin), and SG stands for Stripped Galaxy, i.e. cEs that have a tidal stripping origin. The last column is the resulting outcome of combining all the individual tools together. Green corresponds to properties that have been securely determined (all tools agree or they give complementary information), whereas orange represents the most likely result, as some discriminants did not agree. }
\end{table*}

\subsection{The diverse origins of cEs}
Table 5 summarizes the results discussed above, with the possible origin/evolutionary stage/progenitor type for each cE in the sample considering each diagnostic individually and the final outcome by combining them (last column). For the \textit{no host} galaxies we state only the possible origin, as both an evolutionary stage and a progenitor make no sense under the assumption of an intrinsic origin as low-mass ETGs (LMG). For the rest of the classes, we try to confirm or rule out a stripped origin (SG), in what stage of such process the galaxy is, and finally, what type of progenitor it had most likely. If one of the parameters cannot be determined or is inconclusive from an individual tool, it is left blank (-). Under the stripping scenario, we consider the following evolutionary stages: early, intermediate (int), late and complete. For the progenitor type, E stands for an ETG-like progenitor and S for a spiral-like type. The last column shows the most plausible outcome for each cE combining the information from all the tools. If it is marked in green, it means it is a secure result (all the individual tools agree or complement each other), whereas those in orange show the most likely option when some properties disagree or are inconclusive. It is seen that the use of the different tools in combination is more powerful and can secure the origin for all cEs in the sample, determine about 75\% of their evolutionary stages and provide a plausible progenitor for about 75\% of them. 60\% of the stripped galaxies are more compatible with having ETG-like progenitors, while 30\% would be better represented by a spiral-like type and we cannot determine a plausible progenitor for the remaining 10\%.

\subsubsection{Compact ellipticals with an intrinsic origin} 
Those galaxies that have no identified host nearby, for which the tidal stripping scenario is considered less likely to occur, are typically in the field, although we find two of them in the outskirts of clusters. They show mass ratios close to unity and tend to follow the local scaling relations, except for J16+14 (see below). They have old to intermediate stellar ages ($<t>$= 9.0\,Gyr) and moderately low metallicities ($<$[Z/H]$>$= $-$0.25\,dex) that are compatible with the mass-metallicity relation. Their SFHs are extended with low star forming rates that started at most 12\,Gyr ago, showing long formation timescales compatible with low mass ETGs (\citealt{Thomas2005}; \citealt{Ferre-Mateu2013}). Although some of the individual discriminants are inconclusive, once combined they point towards an intrinsic origin for AHcE0, J07+50, CGCG-036-042 and PGC\,012519. 

The latter is an interesting galaxy. It follows the mass-metallicity relation but deviates slightly from the mass-size one, showing a more compact size for its stellar mass. Both its stellar population and SFH show very old stars and it is the more massive of the galaxies with \textit{no host}. It is thus compatible with being a low-mass elliptical but with a special characteristic: it seems to have not suffered any interactions or star formation episodes since its very early formation. Such objects have been nicknamed `relic galaxies', but so far they have been found in the nearby Universe with masses of $\sim$10$^{11}$M$_{\odot}$ (e.g. \citealt{Trujillo2014}; \citealt{Ferre-Mateu2015}; \citealt{Ferre-Mateu2017}; \citealt{Yildirim2017}). The fact that this galaxy has \textit{no host} but is in a cluster, further reinforces this idea, as such relics seem to prefer the centers of clusters (e.g. \citealt{Poggianti2013}; \citealt{Damjanov2014}; \citealt{PeraltadeArriba2016}). Therefore, we might have found the very first intermediate-mass $\sim$10$^{10}$M$_{\odot}$ relic galaxy.

\subsubsection{Compact Ellipticals caught in the act of stripping} 
While the tidal stripping origin is clear for the galaxies currently being \textit{disrupted}, it is not straightforward to determine in which stage of the process they are. If they just started being stripped by their host, they will still have larger stellar masses and sizes, without departing dramatically from the local scaling relations. They thus still mainly represent the progenitor galaxy and have yet to achieve their final size and mass. We find that this is the case for CZcE44, CZcE57, CZcE95 and NGC\,5567-cE2, which show typically old ages ($<t>$= 11.0\,Gyr) and sub-solar metallicities ($<$[Z/H]$>$ = $-$0.18\,dex). Their early peaked SFHs with subsequent star forming episodes at a very low rate are compatible with having a progenitor of the ETG-like type. Also in an apparently early stage of stripping is NGC\,5567-cE1, although it shows younger ages and more extended SFHs. As it falls in the same age-metallicity area to the three confirmed cEs with and intrinsic origin, we hypothesize that it is a low-mass ETG that has recently started being stripped. Furthermore, we also find that two of the galaxies that were classified as \textit{within halo} (PGC\,038205 and NGC\,2892-cE) are indeed ETG-like galaxies that are about to be stripped, as they show properties that closely follow those seen for \textit{disrupted} cEs at early stages, with no significant deviations form the local scaling relations. 

Later stages in the stripping process can be also identified for a few galaxies, as they strongly deviate from the mass-metallicity relation. They tend to have younger ages ($<t>$= 3.5\,Gyr) and the highest metallicities in the sample ($<$[Z/H]$>$ = 0.0\,dex). Such young ages can be understood from the inferred SFHs, which show an early episode of star formation followed by a second recent burst. This indicates that cold gas from the progenitor is creating new stars when the stripping event happens, and thus the progenitor would be of the spiral-like type. According to the mass-metallicity relation, the progenitor would initially have stellar masses of $\sim$\,5\,$\times$10$^{10}$M$_{\odot}$ for NGC\,2970 and VCC165-cE, and $\sim$10$^{11}$M$_{\odot}$ for PGC\,050564-cE2, AHcE1, AHcE2. CZcE194 is the only galaxy with an unclear evolutionary stage, as some properties suggest a late stage while others to a rather early one. 

\subsubsection{Compact Ellipticals as the end product of the stripping process}
Our findings show that all the galaxies that were initially classified as \textit{near host} are compatible with being the completed stages of tidal stripping. Therefore, they represent the remnant core of the progenitor, with basically very old ages but metallicities that deviate from the mass-metallicity relation ($<t>$= 12.0\,Gyr  and $<$[Z/H]$>$ = $-$0.10\,dex). Such metallicities are indicative of ETG-like progenitors with initial stellar masses of at least 5\,$\times$10$^{10}$M$_{\odot}$. We have thus confirmed such an origin for LEDA\,4544863, NGC\,2832-cE, NGC\,3842-cE and NGC\,5846-cE. Both LEDA\,3126625 and Perseus-cE1 progenitors are more likely to be of the spiral type. NGC\,1272-cE1 (\textit{within halo}) is also, without doubt, the end product of tidal stripping. Additionally, we found that the properties of J16+14, a galaxy without a host, are more compatible with having such stripped origin. This means that  this galaxy could have been first stripped in a more dense environment and it was later ejected from it \citep{Chilingarian2015}.  \\

We are aware that our sample is biased towards objects that are visually classified as \textit{disrupted}, but the number of cEs with \textit{no host} in all known catalogs is very low compared to those that have an associated host. We find that $\sim$85\% of the galaxies in our sample have changed their morphological and stellar properties at some point of their lives via stripping events. The progenitors are typically compatible with having been 10$^{10}$-\,10$^{11}$M$_{\odot}$ galaxies (60\% ETGs/40\% spirals). Only four objects are compatible with being intrinsically low-mass ETGs. Therefore our results seem to favor the stripping mechanism for those galaxies located populous environments such groups and clusters and that have a plausible host nearby, but favors an intrinsic origin for those without a plausible host in looser environments. The finding of cEs that have just started being stripped further emphasizes the fact the even a galaxy with an intrinsic origin might be affected by a stripping process at some point of their life, like the low mass ETG NGC5567-cE1, which has just started its journey of being stripped. 

Determining the origin, evolutionary stage and the progenitor galaxy for the family of cEs is thus a very complex task, with both nature and nurture having a strong role. However, we have shown that it is feasible if combining the information from different key properties. Our sample represents only a fraction of the known cEs to date (about $\sim$1/8) and thus the only way to move forward to disentangle this puzzle is to study statistically complete samples of cEs, but also to extend such study to the low-mass UCDs and to the most massive compact galaxies, both at low an higher redshifts. Further insight from simulations for the formation and evolution of compact objects at all stellar masses is also required to correctly interpret the variety of mechanisms we find observationally.

\section{Summary}
Compact elliptical galaxies are a rare type of compact stellar systems, whose origins are uncertain. They are thought to be either the remnant of a larger and more massive galaxy that has been stripped of its stars or simply the low-mass end of ETGs. In this work we have compiled a representative sample of cEs spanning a range of stellar masses and sizes. They are also representative of the different stages the galaxy will undergo under both possible origins. Galaxies in our sample have been selected from their SDSS images and visually classified into four classes. The sample comprises 11 galaxies that are currently interacting with their host, for which the stripping effect is obvious (\textit{disrupted} class). It also contains galaxies that have an associated host but no signs of interactions are seen. Under the tidal stripping scenario, these cEs could be either the end products of stripping or a system that is about to be stripped. We have visually classified them by \textit{near host} and \textit{within halo}, with 6 and 3 of them, respectively. Finally, we have a sample of 5 galaxies without an associated host (\textit{no host}), which would represent either the unstripped low-mass end of the ETG family or stripped galaxies that escaped from their host.

By studying their structural properties, stellar populations, SFHs, mass ratios, environmental dependencies and how these properties vary through the different classes, we have been able to provide strong constrains on the origin, evolutionary stages and even a possible progenitor, for the majority of cEs in our sample. Our findings are summarized below:
\begin{itemize}
\item The loci the different classes occupy in the mass-size relation provide some hints of the evolutionary stage for a few galaxies only. We see that galaxies more massive than $\sim$10$^{10}$M$_{\odot}$ are typically of the \textit{disrupted} type, with the most massive ones in the sample being at an early stage of tidal stripping. The lowest masses and sizes are instead represented by the \textit{near host} class, which is compatible with the assumption of being the ending product of stripping. The \textit{no host} and the \textit{within halo} classes are located at intermediate masses and sizes, thus not much can be said from this discriminant.
\item cEs show a variety of ages and metallicities, with moderate $\alpha$-enhancements. Such variety reflects the different evolutionary stages the galaxy goes through and also can provide hints about the type of progenitor. The evolutionary stage can be inferred from the mass-metallicity relation. We find that most of the objects that strongly deviate from such relation are either \textit{disrupted} or associated with their host but without interacting signs. Only a few galaxies follow the mass-metallicity relation, with the most massive ones being \textit{disrupted}, \textit{within halo} and \textit{no host} galaxies. The age-metallicity relation provides further evidence for the progenitor type. Young ages and solar or above metallicities are indicative of late stages in the stripping process of a spiral like progenitor that had a reservoir of cold gas. Older ages and sub-solar metallicities, would rather represent an ETG-like progenitor. The \textit{disrupted} galaxies with such properties are those that do not deviate from the mass-metallicity relation, and therefore are at early stages of the process. Those with \textit{no host} show intermediate to old ages with mildly sub-solar values. They have values within the range seen for the \textit{disrupted} galaxies, making them more compatible with having an intrinsic origin. The only exception is J16+14, which has all the properties compatible with a stripped galaxy that escaped from its original environment. 
\item We find a variety of SFHs, with field galaxies showing more extended formation periods (i.e. longer half-mass timescales) and those in clusters showing earlier and faster timescales. Recent episodes of star formation can be explained due to the presence of cold gas in the progenitor that boosted the formation of new stars during the stripping process, and therefore they are more likely to have had a spiral-like one. Those that show early formation epochs with either no posterior formation events or an extended, low-rate of star formation, would represent a progenitor of the ETG type. SFHs that are very extended in time with intermediate ages are representative of intermediate and low-mass ETGs, as found for most of the \textit{no host} galaxies in our sample. 
\item We find that PGC\,012519 could be the first reported low mass analog of the more massive local relic galaxies. Its properties are compatible with being an intermediate-mass ETG that formed at very early times (given by its SFH, mass and size) and that remained as such since then, frozen over cosmic time. Its location in the outskirts of a cluster further reinforces this hypothesis.
\item Mass ratios are seemingly the least reliable of the tools employed here. They only become useful for those galaxies that show mass ratios larger than 2. These correspond to low mass galaxies in our sample, as previously found for compact systems. However, this tool has allowed us to secure the origin of J16+14, a galaxy classified as \textit{no host} but with most its properties being compatibles with a stripped origin.
\item In general, each discriminant tool alone is not capable of determining the origin, evolutionary stage and progenitor type. Only when we combine the different tools  can we robustly determine such parameters for most of the galaxies in the sample. We find that the majority of the sample (85\%) is or has been stripped at some point, while only 15\% could be considered the very low mass end of the ETG family. We find that roughly 30\% of the cEs are compatible with having a spiral-like progenitor, while 60\% show indications of having an ETG like progenitor of varying stellar mass. Therefore our results suggest that the main mechanism for shaping the population of cEs is the tidal stripping of larger and more massive galaxies, particularly in groups and clusters, whereas galaxies with no nearby host and loose environments seem to rather have an intrinsic origin. Larger, statistically complete samples are needed to address this issue securely. 
\end{itemize}

\section*{Acknowledgements}
We thank the referee for the useful comments provided, which have greatly improved the clarity of the manuscript. AFM and DAF thank the ARC for financial support via DP160101608. AJR was supported by NSF grant AST-1515084 and as a Research Corporation for Science Advancement Cottrell Scholar. We thank Ignacio Mart\'in-Navarro, Anish Seshadri, Carolyn Wang and the students in SJSU ASTR 155 for exploratory work on cEs that inspired this project, and acknowledge Vasily Belokurov for alerting us about the NGC\,2970 stream. We also thank Alister W. Graham for useful insight and comments.



\appendix
\section{Galaxy size estimates}
Some of our objects have been previously studied individually or in larger samples and therefore have published sizes. We first investigate the impact of using SDSS sizes by comparing those with the published ones (Section 3.1). This is seen in Figure 15 (top panel), showing that for only a few galaxies the SDSS is somewhat over estimated. This happens typically for the galaxies in the \textit{disrupted} or \textit{within halo} classes. Such overestimation on the size in SDSS is related to the magnitude that is derived from the assumed model in SDSS. This magnitude is used to determine our new stellar masses in Section 3.2, which we expect to vary too. The lower panel of Figure 15 shows the comparison of previously published stellar masses \textit{versus} our $\mathrm{M_{*,L}}$. It is seen that the objects classified as \textit{within halo} are, again, the ones deviating the most. Therefore, we will use the literature values for both size and stellar masses for NGC\,1272-cE1, NGC\,2892-cE and PGC\,038205.

We then investigate how much the sizes and masses of the galaxies with only SDSS information could vary. From Figure 15 it is inferred that the largest variations occur for galaxies typically larger than $\sim$500\,pc, with a size variation that typically decreases by a factor of $\sim$2 (except for NGC\,2892-cE and PGC\,038205, in which case it is a factor of $\sim$10, as their SDSS size are too large ). Such size variations imply a decrease in the stellar mass of roughly 0.5\,-\,0.7\,dex (except for the mentioned special cases). Figure 16 shows how much we could expect our galaxies with only SDSS values to vary following the values in Figure 15. From our sample of 25 cEs, only 9 have solely SDSS values (crossed red pentagons). From them, only 4 have sizes larger than 500\,pc: CZcE57, CZcE95 and its companion PGC050564-cE2, and NGC5567-cE2. Interestingly, they all belong to the \textit{disrupted} class and are, in fact, the most disrupted cases seen in Figure 5. We thus impose a size variation of $\times$0.5 in size and 0.7\,dex in stellar mass and see how they would move around the mass-size and mass ratio figures, represented by the yellow pentagons. 

It is straightforward to see that nothing changes in the stellar mass \textit{vs} mass ratios panel, as all the crossed objects where already within the intrinsic scatter and remain within it after the variation. In the mass\,-\,size plane, we can see that for the two most massive galaxies in our sample, CZcE57 and NGC5567-cE2, such variations would bring them closer to the bulk of our galaxies, although they would remain to be the most massive ones. The variations for the other two galaxies do not imply any major changes either, as they had been considered to be at an intermediate phase. Therefore, such variations can have an impact on the exact stage of stripping for these four galaxies but we do not expect any other major changes, as discussed in Section 4.1. Therefore we use from now on the SDSS estimates for those galaxies without literature values but we consider them as upper limits, flagging them and treating them cautiously in the discriminant tools where sizes play a relevant role. 

\begin{figure*}
\centering
\includegraphics[scale=0.7]{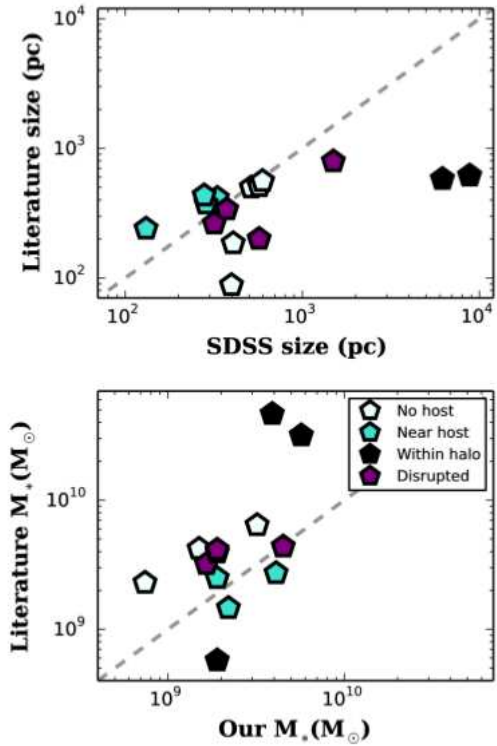}\\
\caption{Size (upper panel) and stellar mass (bottom panel) comparison between SDSS and published values. The galaxies have been coloured according to their class: white for galaxies with \textit{no host}, purple for those being \textit{disrupted}, black for galaxies within the halo of their host and cyan for galaxies near a host. The dashed lines are the one-to-one relations to guide the eye. }
\label{figure:15}
\end{figure*}

\begin{figure*}
\centering
\includegraphics[scale=0.7]{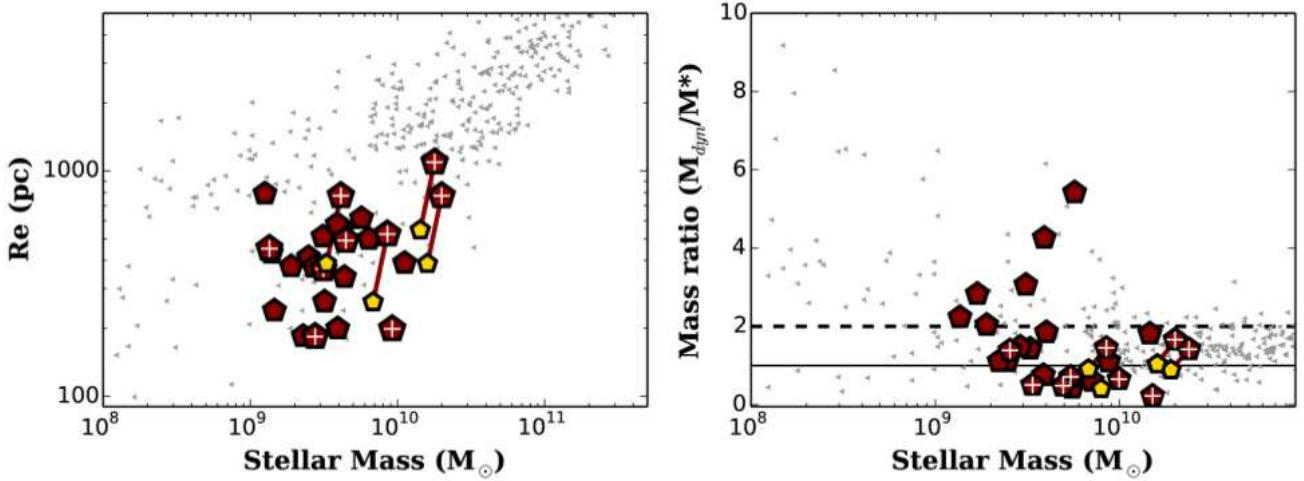}\\
\caption{Size overestimation impact. Left panel reproduces the mass-size relation as in Figure 12 (top) and the right panel reproduces the mass ratios as in Figure 14, with small grey symbols representing the N14 sample. The 25 cEs in our sample are shown with red pentagons, those with a cross to highlight that only a SDSS size is available. For such galaxies, if they are larger than 500\,pc and are embedded within the host halo or in streams, their sizes are most likely overestimated. The yellow pentagons show how much they could change in size ($\sim\,\times$0.5) and stellar mass ($\sim\,\times$0.7\,dex) and the new position they would occupy in both planes. Most importantly, such variations do not change any of the results on the basis of the diverse discriminant tools.}
\label{figure:16}
\end{figure*}

\section{Classical index approach}
\begin{figure*}
\centering
\includegraphics[scale=0.70]{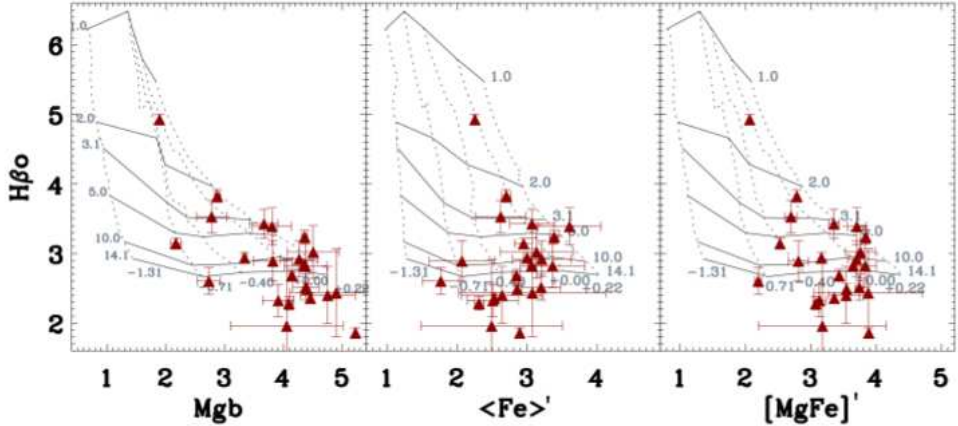}\\
\caption{Model grids of the different index-index pairs used. Left and middle panels show the age-sensitive H$\beta_{o}$ index against the metal indicators Mgb$_{5177}$ and  <Fe>$^{\prime}$, which are used to derive the $\alpha$-enhancement ratios. The right panel shows the H$\beta_{o}$ - [MgFe]$^{\prime}$ pair, which is used to derive the SSP ages and metallicities shown in Table  3.}
\label{figure:17}
\end{figure*}

\begin{figure*}
\centering
\includegraphics[scale=0.50]{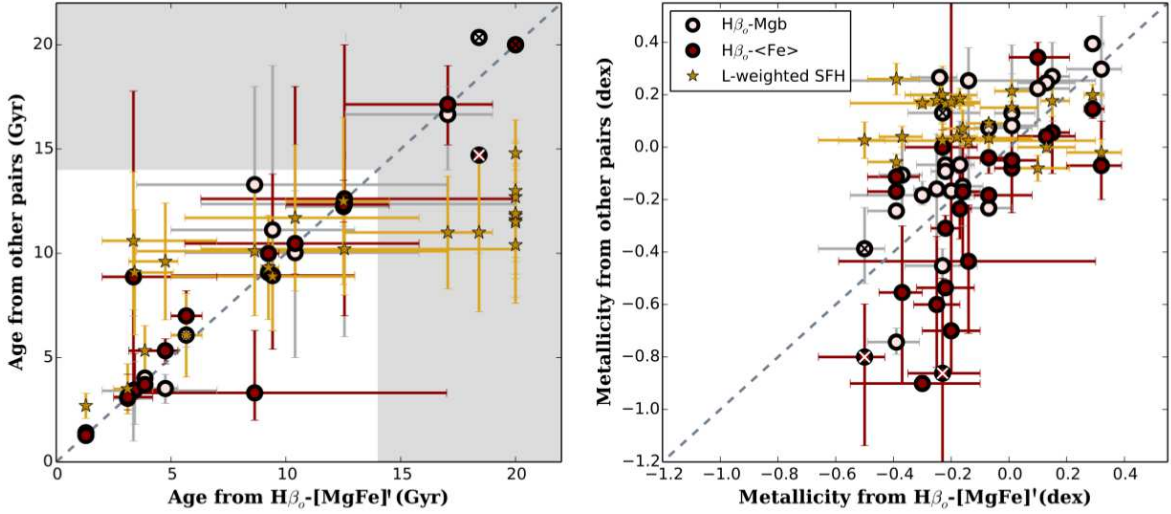}\\
\caption{Comparison of the age and metallicity obtained from different index-index pairs and the full-spectral fitting (luminosity-weighted). The estimates used throughout this paper from the H$\beta_{o}$ - [MgFe]$^{\prime}$ index are compared  to the other pairs, showing good agreement in both the ages, and metallicities. Galaxies with a cross correspond to the low S/N ones. The dashed line is the one-to-one relation to guide the eye and the shaded area in the age panel represents the extrapolated age values from the index-index grids. Any galaxy that falls in that region should be considered as old as the models employed ($t\sim$14\,Gyr).}
\label{figure:18}
\end{figure*}

\begin{figure*}
\centering
\includegraphics[scale=0.40]{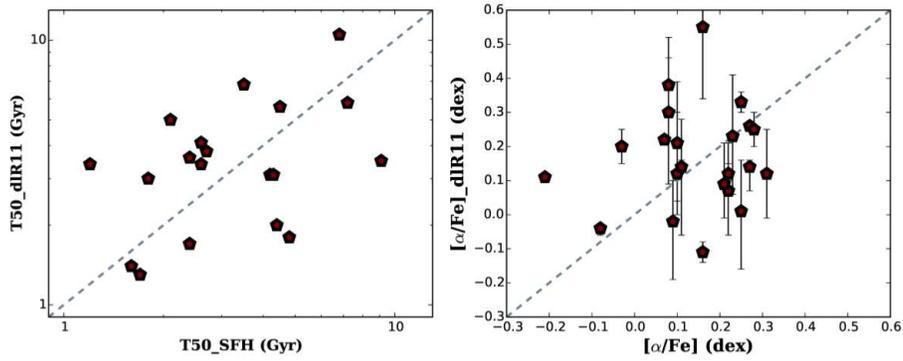}
\caption{Left panel shows the T50 values obtained from the star formation histories compared to those inferred from the empirical relation of \citet{delaRosa2011}. The right panel shows the [$\alpha$/Fe] ratios measured from the line indices as described above, compared to those derived with the same empirical relation of \citet{delaRosa2011} but using the T50 obtained from the SFHs. The dashed line is the one-to-one relation to guide the eye.}
\label{figure:19}
\end{figure*}

\begin{figure*}
\centering
\includegraphics[scale=0.53]{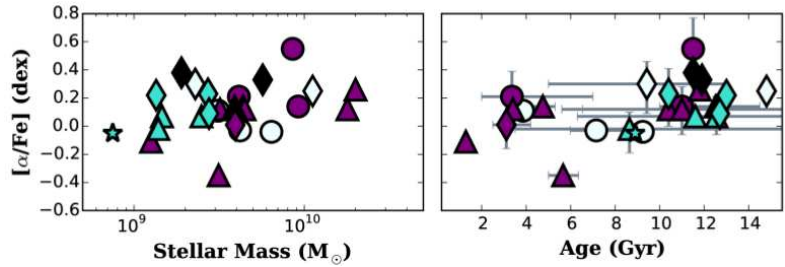}\\
\caption{Scaling relations of compact systems with the $\alpha$-enhancement. Left panel correspond to the relation with stellar mass and right one with the line index ages. The prototype cE M32 is shown in all panels as a cyan star for comparison.}
\label{figure:20}
\end{figure*}

\begin{figure*}
\centering
\includegraphics[scale=0.40]{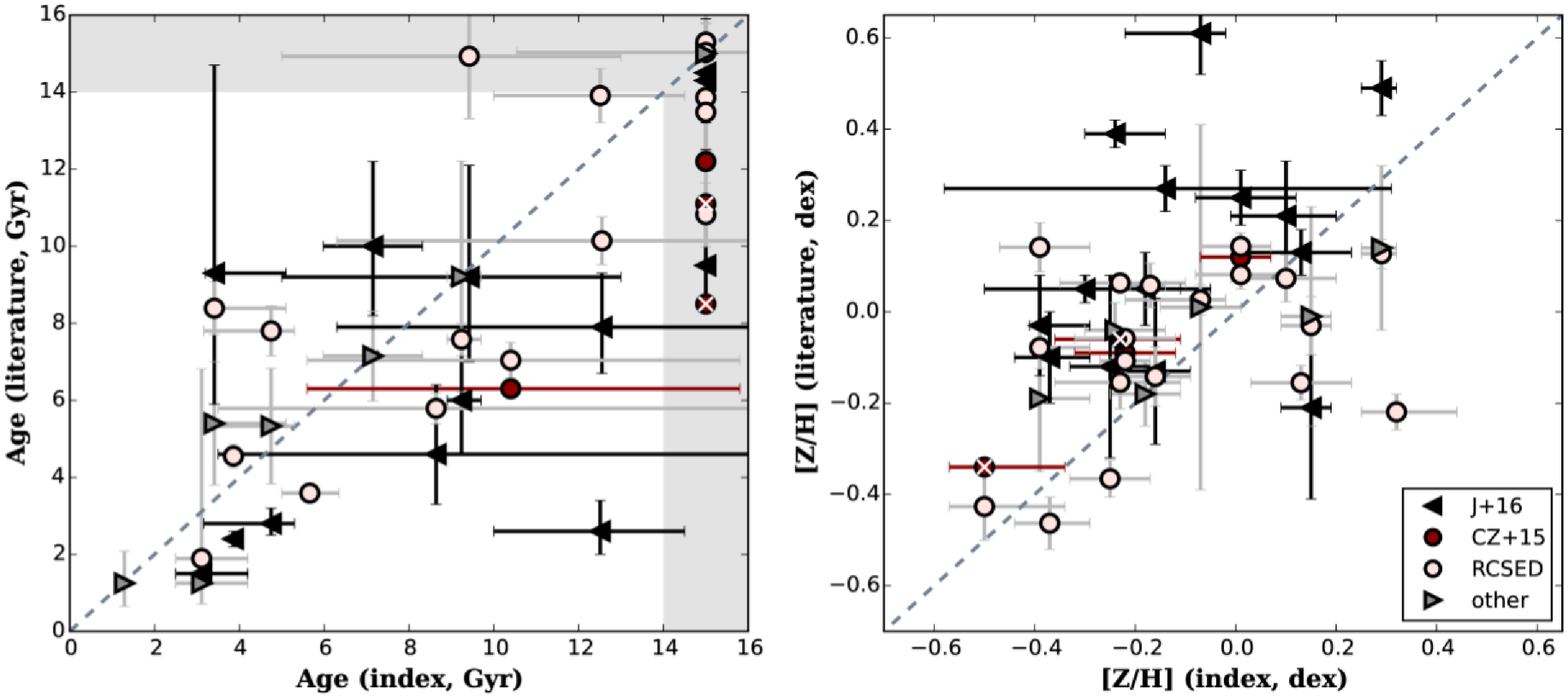}\\
\caption{Our stellar population parameters compared to previously published values (from the line index approach). As in Figure 18, galaxies falling in the shaded area have been extrapolated due to being outside the model grids, thus any galaxy in that area is treated as if its age was $t\sim$14\,Gyr. The dashed line is the one-to-one relation to guide the eye. Our values show good agreement with the samples of CZ+15 and RCSED, but present a larger scatter from those in J+16, which could be attributed to aperture effects.}
\label{figure:21}
\end{figure*}

In this section we present in detail the stellar populations derived from the classical line index approach. We measure the most relevant indices in our spectral region, but we focus on H$\beta_{o}$ \citep{Cervantes2009} as our main age-sensitive index and Mgb$_{5177}$, Fe5270 and Fe5335 as the metallicity-sensitive ones. We use the same SSP models as in section 3.2 to create index-index model grids to compare with the line index measurements, from which we can infer a mean age and metallicity for each pair. We obtain three different pairs of age-metallicity estimates from the grids in Figure 17, which compare H$\beta_{o}$ with Mgb (left), <Fe>$^{\prime}$ (middle) and [MgFe]$^{\prime}$, (right), where the last two are the following composite indices:\\
<Fe>$^{\prime}$ =\,(0.72 $\times$ Fe5270 + 0.28 $\times$ Fe5335); [Mg/Fe]$^{\prime}$= $\sqrt{\mathrm{Mgb \,\,\times <Fe>^{\prime}}}$; \citep{Thomas2003}.

Figure 18 shows the ages and metallicities derived from different pairs from Figure 17 and the luminosity-weighted estimate from the full-spectral-fitting. There is a remarkable agreement between the age estimates from all indices, but also for the [M/H] values. When we refer to index estimates, e.g. M$_{*,ind}$ in the previous sections, we will use the age and metallicity from the H$\beta_{o}$ - [MgFe]$^{\prime}$ pair. This pair has been shown to give the best approximation for the total metallicity ([Z/H]), as [MgFe]$^{\prime}$ is not affected by $\alpha$ enhancements.

We obtain the [$\alpha$/Fe] ratios using the same approach as in \citet{Vazdekis2015}. The model grids obtained when plotting the age sensitive index H$\beta_{o}$ against the metallicity sensitive indices Mgb and <Fe>$^{\prime}$, render two metallicity estimates, $\mathrm{Z_{Mgb}}$ and $\mathrm{Z_{Fe}}$ respectively. The proxy for [$\alpha$/Fe] is then calculated as [$\mathrm{Z_{Mg}/Z_{Fe}}$] = $\mathrm{Z_{Mgb}\,-\,Z_{Fe}}$ and using the empirical relation of \citet{Vazdekis2016} we obtain the real ratio with [$\alpha$/Fe] = 0.59 $\times$ [$\mathrm{Z_{Mg}/Z_{Fe}}$].

With these [$\alpha$/Fe] ratios, we have derived T50 values using the empirical formula of \citet{delaRosa2011}. To be consistent within our analysis, we compare those T50s with the ones inferred from the full-spectral fitting in Section 3.3. The good agreement between the two estimates, as shown in Figure 19, allows us to derive [$\alpha$/Fe] values for those galaxies where the index measurement did not provide a reliable result (marked with a \ddag  \,in Table 3). 

Figure 20 shows the scaling relations for the stellar mass and stellar ages, similar to Figure 12 but with the $\alpha$-enhancements this time. We see that the correlation is very mild, with older galaxies showing higher abundance ratios, but with no specific trend with the stellar mass.

We finally compare the results of our stellar population analysis with those previously published. Figure 21 shows that our results are in good agreement with CZ+15. However, there is a wider spread on the agreement with the values published in J+16. Those galaxies that also have other published values, e.g. AHcE0, AHcE1 and AHcE2 (\citealt{Huxor2011}; \citealt{Huxor2013}) are more consistent with our new measurements and with the luminosity-weighted full-spectral fitting approach. All these differences, which are the greatest in the metallicity values, can be associated with using different SSP models (which are known to have a bigger impact on the total metallicities than on the ages) and a different spatial coverage of the galaxy (SDSS 3" fiber \textit{vs} longslit).
\bibliographystyle{apj}
\bibliography{sdss_ce_astroph}

\end{document}